\documentclass[prd,aps,
floats,letterpaper,floatfix,
groupedaddress,superscriptaddress
,nofootinbib
,twocolumn
]{revtex4}
\usepackage{amsmath}
\usepackage{amsfonts}
\usepackage{bm}
\usepackage{graphicx}
\usepackage[usenames]{color}

  \newcommand{\mnras}{Mon.\ Not.\ R.\ Astron.\ Soc.}

   \newcommand{\aap}{Astron.\ Astrophys.}
   
   \newcommand{\aj}{Astron.\ J.}

\allowdisplaybreaks

\begin{document}

\title{Higher-order effects in the dynamics of hierarchical triple systems. II.  Second-order and dotriacontapole-order effects}
\author{
Landen Conway} \email{conwayl@ufl.edu}
\affiliation{Department of Physics, University of Florida, Gainesville, Florida 32611, USA}

\author{
Clifford M.~Will} \email{cmw@phys.ufl.edu}
\affiliation{Department of Physics, University of Florida, Gainesville, Florida 32611, USA}
\affiliation{GReCO, Institut d'Astrophysique de Paris, CNRS,\\ 
Sorbonne Universit\'e, 98 bis Bd.\ Arago, 75014 Paris, France}

\date{\today}

\begin{abstract}
We analyze the long-term evolution of hierarchical triple systems in Newtonian gravity to second order in the fundamental quadrupolar perturbation parameter, and to sixth order in $\epsilon = a/A$, the ratio of the semimajor axes of the inner and outer orbits.   We apply the ``two-timescale'' method from applied mathematics to the Lagrange Planetary Equations for the inner and outer orbits, in which each osculating orbit element is split into an orbit averaged part that evolves on the long perturbative timescale, and an ``average-free'' part that is oscillatory in the orbital timescales.  Averages over the two orbital timescales are performed using the well-known ``secular approximation'', carefully adapted to account for the time integrals that produce the oscillatory solutions.  We also incorporate perturbative corrections to the relation between time and the orbital phases, or ``anomalies''.  We place no restrictions on the masses of the three bodies, on the relative orbit inclinations or on the eccentricities, beyond the requirement that the quadrupolar perturbation parameter and $\epsilon$ both be ``small''.  The result is a complete set of long-timescale evolution equations for the averaged elements of the inner and outer orbits.  At first order in perturbation theory, we obtain the dotriacontapole contributions explicitly at order $\epsilon^6$, augmenting earlier well-known results at quadrupole, octupole and hexadecapole orders.  At second order in perturbation theory, i.e. quadratic in the quadrupole perturbation amplitude, we find contributions that scale as $\epsilon^{9/2}$ (found in earlier work), $\epsilon^{5}$, $\epsilon^{11/2}$, and $\epsilon^{6}$.  At first perturbative order and dotriacontapole order, the two averaged semimajor axes are constant in time (and we prove that this holds to arbitrary multipole orders); but at second perturbative order, beginning at $O( \epsilon^{5})$, they are no longer constant.  Nevertheless we verify that the total averaged energy of the system is conserved, and we argue that this behavior is not incompatible with classical theorems on secular evolution of the semimajor axes.

\end{abstract}

\pacs{}
\maketitle

\section{Introduction}
\label{sec:intro}
Three-body systems are common in our universe.  As many as a third of the stars in our Galaxy could be in triple systems.  Closer to home, the Earth-Moon-Sun system is a triple system that confounded scientists from Newton to Hill.  Three-body interactions play a role in the evolution of larger systems such as globular clusters.  Numerous exoplanets exist in systems containing two stars.  The rich dynamics of triple systems have been exploited in situations relevant to general relativity, such as providing possible pathways to compact binary inspiral and merger (for a recent textbook, see, e.g. \cite{2006tbp..book.....V}).

Three-body dynamics are inherently unstable, with two of the bodies tending to form a tight binary, while the third body is flung to large distances, orbiting the inner pair so as to form an ``outer'' binary.   Such systems are called ``hierarchical triples''.  To a first approximation, each binary evolves on a Keplerian orbit, with orbital elements that are constant in time.    Including the effects of all the mutual gravitational interactions results in dynamics that can take place on timescales much longer than that of the two orbital periods.  Under suitable constraints placed on the masses and separations of the two binaries, the dynamics of this system can be studied perturbatively. Perturbative studies of hierarchical three-body systems focus on these long timescale (or secular) effects.

Treating the gravitational interactions as a perturbation of the Keplerian behavior of the two orbits, and expanding them in a sequence of multipoles, one finds that the lowest order non-trivial result comes from the ``quadrupole'' part of the perturbation, where the secular evolution is found by performing a time average over the two orbital timescales.  Working to this order and with a circular outer binary, Kozai and Lidov \cite{1962AJ.....67..591K,1962P&SS....9..719L} found the notable result of large periodic variations in eccentricity and inclination of the inner orbit, which is now known colloquially as the Kozai-Lidov effect (with the name von Zeipel often added in recognition of his insights). This result has seen widespread application in an array of astrophysical situations including solar system satellites and asteroids, compact binary coalescence, and exoplanet dynamics. 
	
Generalizing this result to include eccentric outer orbits, and including successive multipolar perturbation to the equations of motion, Naoz et al \cite{2011Natur.473..187N,2013MNRAS.431.2155N} 
calculated the secular equations of motion through octupole order (for earlier work, see
\cite{1999MNRAS.304..720K,2000ApJ...535..385F,2002ApJ...578..775B}). The incorporation of octupole order effects allows for complete ``flips'' of the inner orbital plane accompanied by  excursions of the eccentricity to nearly unity.  
Hexadecapole effects were calculated and studied by Will \cite{2017PhRvD..96b3017W}  and were shown to suppress orbital flips under some circumstances and to incite them under others.  

At the next multipole order, known as dotriacontapole order, the effects have the same scaling in terms of the small ratio of semimajor axes of the inner and outer orbits ($a$ and $A$ respectively) as do the ``square'' of quadrupole order effects.  Such ``quadrupole squared'' effects would occur at second order in perturbation theory, induced by the ``feedback'' of first order perturbations into the equations of evolution of the orbit elements.  The dominant effects of these terms were explored in \cite{2021PhRvD.103f3003W} (hereafter denoted Paper I).  These effects were also explored using a vectorial approach in 
\cite{2016MNRAS.458.3060L} and using an effective field theory (EFT) approach in
\cite{2021PhRvD.104b4016K,2023PhRvD.107d4011K}; all results were in complete agreement.   

The surprising discovery was that, while first-order perturbation theory (achieved by holding all orbit elements fixed while performing the orbital averages) leads to a regular sequence of multipolar perturbations with amplitudes $(a/A)^3$ for quadrupole terms, $(a/A)^4$ for octupole terms,  $(a/A)^5$ for hexadecapole terms, and $(a/A)^6$ for dotriacontapole terms, the dominant ``quadrupole squared'' terms had amplitudes $(a/A)^{9/2}$, larger than expected by a factor of the ratio of the outer orbital period to the inner orbital period.  We denote these as ``dominant'' contributions, while the accompanying  ``sub-dominant'' contributions are expected to have amplitudes scaling as $(a/A)^6$.
Tremaine \cite{2023MNRAS.522..937T} provides a thorough examination of this effect and compares approaches by different authors. 
 
In this paper we endeavor to complete the work that motivated Paper I, namely to compute the first-order dotriacontapole terms and the sub-dominant quadrupole-squared terms, both of which scale as $(a/A)^6$.  However, in the course of doing this, we discovered a plethora of additional second-order contributions to the evolutions, with scalings of $(a/A)^5$ (the same as first-order hexadecapole),  $(a/A)^{11/2}$, and $(a/A)^6$ (the same as first-order dotriacontapole).   These new effects arise from  dominant quadrupole-squared contributions from the second-order feedback of variations of {\em outer} orbit elements;  dominant contributions from ``quadrupole-octupole'' cross terms (with both inner and outer orbit element feedback); and contributions from corrections to the relationship between time and the orbital phases.
 
Our results contained an additional  surprise.
Until now, all results calculated in the literature at first-order in perturbation theory, at quadrupole, octupole, and hexadecapole orders (as well as at the dotriacontapole order calculated in this paper), have the property that the semimajor axes of both orbits show no long timescale variation; in other words, after averaging over the orbital timescales, $da/dt$ and $dA/dt$ are zero.   We provide a simple proof of this to arbitrary multipole orders in  Appendix \ref{app:semimajor}.   However, at second order in perturbation theory, this is no longer true in general (the dominant quadrupole-squared terms calculated in Paper I do keep $a$ and $A$ constant).   These variations in $a$ and $A$ first arise at order $(a/A)^5$, the same as hexadecapole order; these terms arise from the feedback of quadrupole perturbations of outer orbit elements into the quadrupole perturbations, combined with corrections to the relation between time and the outer true anomaly.  Quadrupole-octupole cross terms also lead to variations in $a$ and $A$.  We believe this to be the first explicit calculation of second-order effects in hierarchical triples in pure Newtonian theory that display nonzero variations of inner and outer semi-major axes.  (Such variations do arise when post-Newtonian-Quadrupole ``cross-term'' effects are included \cite{2014PhRvD..89d4043W,2020PhRvD.102f4033L}).

This seems to fly in the face of conventional wisdom, which holds that the semimajor axes of hierarchical triples must be constant, as proven by Poisson, Poincar\'e, Tisserand and others \cite{1897BuAsI..14..241P,Tisserand89,1902AnPar..23A...1A,1978A&A....68..199D}
(for a pedagogical treatment, see Sec. 8.7 of \cite{HagiharaVol2}).   However these theorems prove only that the semimajor axes contain no ``purely secular'' terms, i.e. terms proportional to $t$.  We will see that our long-term variations in $a$ and $A$ are purely {\em periodic} in linear combinations of the two pericenter angles, fully consistent with the theorems.

Furthermore, the variations we find do {\em not} imply a failure of energy conservation: in fact for all of the effects we have calculated, we verify that the {\em total} energy of the system, containing both Keplerian contributions (proportional to $1/a$ and $1/A$) and interaction terms, is constant.  But our results do suggest that, at second order,
 the Keplerian energies of both the inner and outer orbits are no longer separately constant.

The remainder of this paper presents details. In Sec.\ II we set up the problem, define the orbit elements, describe the ``Lagrange planetary equations'' which determine their evolution, expanded to dotriacontapole order, and discuss the conserved quantities. In Sec.\ III we review the ``two-timescale method'' for solving the planetary equations as a problem in perturbation theory, and review how the averaging over the two orbital timescales is carried out.   Section IV describes our qualitative expectations for the second-order contributions.  This is followed by concluding remarks in Sec.\ V. In Appendix \ref{app:results} we provide detailed expressions for the calculated secular equations; in Appendices \ref{app:Y} and \ref{app:timeconversion} we provide details of some specific derivations; and in Appendix \ref{app:semimajor} we display a simple proof to all multipole orders of the secular constancy of the semimajor axes in first-order perturbation theory.

\section{Hierarchical triple systems and the Lagrange planetary equations}
\label{sec:triples}

\begin{figure}[t]
\begin{center}

\includegraphics[width=3.4in]{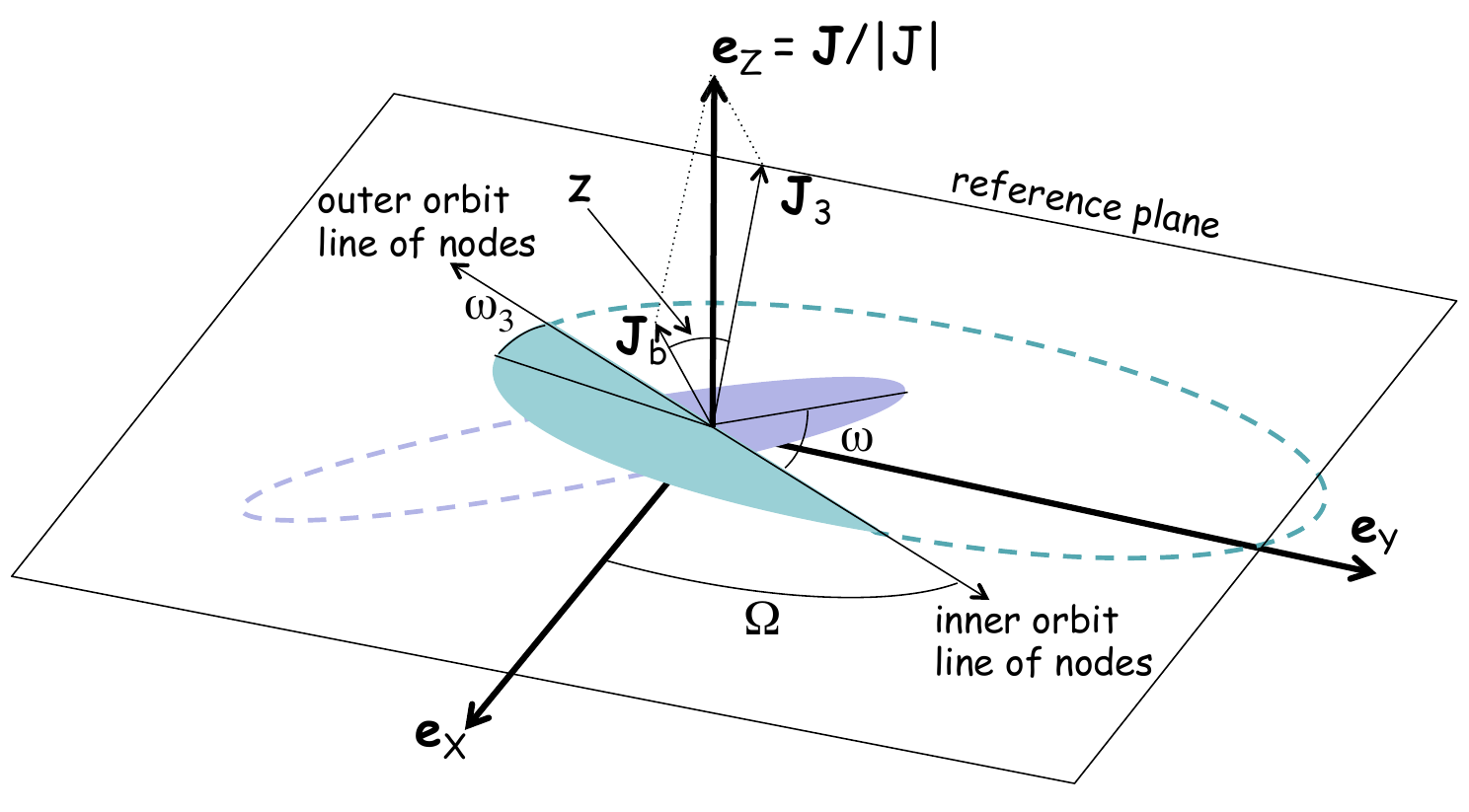}

\caption{Orientation of inner and outer orbits. (Color figures in online version.)
\label{fig:orbits} }
\end{center}
\end{figure}

We consider a  three-body system shown in Fig.\ \ref{fig:orbits}, with bodies 1 and 2 making up the ``inner'' binary, and with body 3 chosen as the outer perturbing body.  The orbital separation of the inner binary ($r \equiv |{\bm x}| = |{\bm x}_1 - {\bm x}_2|$)  is assumed to be small compared to the distance of the outer binary from the inner binary's center of mass  ($R \equiv |{\bm X}| = |{\bm x}_3 - {\bm x}_{\rm cm}|$). The equations of motion will be expanded in terms of that small ratio.   We define $m \equiv m_1 + m_2$, $M \equiv m + m_3$, $\eta \equiv m_1m_2/m^2$ and $\eta_3 \equiv m_3 m/M^2$, with the convention that $m_1 \le m_2$ .   

The equations of motion then have the general form
\begin{align}
a^j &= - \frac{Gm n^j}{r^2} + \frac{Gm_3 }{R^2} \sum_{\ell = 1}^\infty \frac{(2\ell +1)!!}{\ell !} \left ( \frac{r}{R} \right )^\ell  g_\ell
n^L N^{\langle jL \rangle}  \,,
\nonumber \\
A^j &=-  \frac{GM N^j}{R^2} -  \eta\frac{GMr}{R^3} \sum_{\ell = 1}^\infty \frac{(2\ell +3)!!}{(\ell+1) !} \left ( \frac{r}{R} \right )^\ell g_\ell
\nonumber \\
& \qquad \times
n^{L+1} N^{\langle j(L+1) \rangle} \,,
\label{eq2:eom2}
\end{align}
where $\bm{a} \equiv d^2 \bm{x}/dt^2$ and $\bm{A} \equiv d^2 \bm{X}/dt^2$,  $G$ is Newton's constant, and $g_\ell \equiv (m_2/m)^\ell - (-m_1/m)^\ell$.   The superscript $\langle \dots \rangle$ denotes a symmetric tracefree product of the unit vectors (for a review see \cite{PW2014}).   
Note that the perturbing terms in the equation for $A^j$ depend on the inner binary's reduced mass parameter $\eta$; this is to be expected, since in the limit in which body 1 is a test body, $\eta \to 0$, and the third body moves on an unperturbed Keplerian orbit around the massive body 2. 

The various multipolar orders are denoted quadrupole ($\ell=1$), octupole ($\ell=2$), hexadecapole ($\ell=3$), dotriacontapole ($\ell=4$), and so on.

We define the osculating orbit elements of the inner and outer orbits in the standard manner: 
\begin{eqnarray}
r &\equiv& p/(1+e \cos f) \,,
\nonumber \\
{\bm x} &\equiv& r {\bm n} \,,
\nonumber \\
{\bm n} &\equiv& \left [ \cos \Omega \cos(\omega + f) - \cos \iota \sin \Omega \sin (\omega + f) \right ] {\bm e}_X 
\nonumber \\
&&
 + \left [ \sin \Omega \cos (\omega + f) + \cos \iota \cos \Omega \sin(\omega + f) \right ]{\bm e}_Y
\nonumber \\
&&
+ \sin \iota \sin(\omega + f) {\bm e}_Z \,,
\nonumber \\
{\bm \lambda} &\equiv& d{\bm n}/df \,, \quad \hat{\bm h}={\bm n} \times {\bm \lambda} \,,
\nonumber \\
{\bm h} &\equiv& {\bm x} \times {\bm v} \equiv \sqrt{Gmp} \, \bm{\hat{h}} \,,
\label{eq2:keplerorbit1}
\end{eqnarray}
where (${\bm e}_X,\,{\bm e}_Y ,\,{\bm e}_Z$) define a reference basis, with ${\bm e}_Z$ aligned along the total angular momentum of the system, and with the ascending node of the inner orbit oriented at an angle $\Omega$ from the $X$-axis, as shown in Fig.\ \ref{fig:orbits}.  From the given definitions, it is evident that ${\bm v} = \dot{r} {\bm n} + (h/r) {\bm \lambda}$ and $\dot{r} = (he/p) \sin f$. 

The outer orbit is defined in the same manner, with orbit elements $P$, $E$, $\omega_3$, $\Omega_3$, and $\iota_3$ replacing $p$, $e$, $\omega$, $\Omega$ and $\iota$, $\bm{\Lambda}$ and $\bm{H}$ replacing $\bm{\lambda}$ and $\bm{h}$, and $F$ replacing $f$.  
The semimajor axes of the two orbits are defined by $a \equiv p/(1-e^2)$ and $A \equiv P/(1-E^2)$.

We then write down the ``Lagrange planetary equations'' for the evolution of the inner orbit elements,
\begin{eqnarray}
\frac{dp}{dt} &=& 2 \sqrt{\frac{p^3}{Gm}} \frac{{\cal S}}{1+e \cos f} \,,
\nonumber \\
\frac{de}{dt} &=& \sqrt{\frac{p}{Gm}} \left [ \sin f \, {\cal R} + \frac{2\cos f + e +e\cos^2 f}{1+ e\cos f} {\cal S} \right ]\,,
\nonumber \\
\frac{d\varpi}{dt} &=& \frac{1}{e}\sqrt{\frac{p}{Gm}} \left [ -\cos f \, {\cal R} + \frac{2 + e\cos f}{1+ e\cos f}\sin f {\cal S} 
\right ] \,,
\nonumber \\
\frac{d\iota}{dt} &=& \sqrt{\frac{p}{Gm}} \frac{\cos (\omega +f)}{1+ e\cos f} {\cal W} \,,
\nonumber \\
 \frac{d\Omega}{dt} &=& \sqrt{\frac{p}{Gm}} \frac{\sin (\omega +f)}{1+ e\cos f} \frac{\cal W}{\sin \iota}\,,
\label{eq2:lagrange}
\end{eqnarray}
where ${\cal R}$,
 ${\cal S} $ and
 ${\cal W} $ are the radial, azimuthal and out-of-plane components of the perturbing accelerations from Eqs.\ (\ref{eq2:eom2}).
The auxiliary variable $\varpi$ is defined such that the change in pericenter angle is given by $\dot{\omega} = \dot{\varpi} -  \dot{\Omega} \cos \iota$.  These are augmented by a ``sixth'' planetary equation that relates $f$ to time via
\begin{align}
\frac{df}{dt} &= \sqrt{\frac{Gm}{p^3}} (1+e\cos f)^2 
\nonumber \\
& \quad 
+ \frac{1}{e} \sqrt{\frac{p}{GM}} \left [ \cos f \, {\cal R} - \frac{2+e\cos f}{1+e \cos f} \sin f \, {\cal S} \right ] 
\nonumber \\
&= \left (\frac{df}{dt} \right )_K - \frac{d\varpi}{dt}\,,
\label{eq:dfdt}
\end{align}
where $(df/dt)_K$ denotes the Keplerian expression. 

The planetary equations for the outer binary take the form of Eqs.\ (\ref{eq2:lagrange}) and (\ref{eq:dfdt}), with suitable replacements of all the relevant variables, and perturbation components ${\cal R}_3$,
 ${\cal S}_3$ and ${\cal W}_3$.  These equations are an exact reformulation of the six second-order differential equations (\ref{eq2:eom2}) as 12 first-order differential equations for the orbit elements, $f$ and $F$.

For purely technical reasons, we will actually use the ``eccentric anomaly'' $u$ to describe the inner orbit, rather than the true anomaly $f$.  This allows us to avoid having to evaluate integrals involving negative powers of $(1+ e \cos f)$.  In terms of $u$, we have 
\begin{align}
r &=a (1-e \cos u) \,,
\nonumber \\
\tan \frac{1}{2} f &= \sqrt{\frac{1+e}{1-e}} \tan \frac{1}{2} u \,.
\label{eq:radiusu}
\end{align}
The relationship between $u$ and time $t$ is given by
\begin{align}
\frac{du}{dt} & = \left (\frac{du}{dt} \right )_K
 - \frac{1-e\cos u}{\sqrt{1-e^2}} \frac{d\varpi}{dt} - \frac{\sin u}{1-e^2} \frac{de}{dt} \,,
 \label{eq:dudt1}
\end{align}
where
\begin{equation}
\left (\frac{du}{dt} \right )_K =\sqrt{\frac{Gm}{a^3}} (1-e \cos u)^{-1}  \,.
 \label{eq:dudtK}
\end{equation}

The Newtonian three-body equations of motion admit a conservation law for total angular momentum, $d{\bm J}/dt = 0$, both for the exact equations and for the multipole expansion, order by order.
Combining the Lagrange planetary  equations and inserting the perturbing accelerations at arbitrary multipole order, it is straightforward to verify directly that
\begin{align}
\frac{d}{dt}  & \left ( \Omega - \Omega_3 \right ) = 0 \,,
\nonumber \\
\frac{d}{dt} &\left ( m\eta \sqrt{Gmp} \sin \iota - M\eta_3 \sqrt{GMP} \sin \iota_3 \right ) = 0 \,,
\nonumber \\
\frac{d}{dt} &\left ( m\eta \sqrt{Gmp} \cos \iota + M\eta_3 \sqrt{GMP} \cos \iota_3 \right ) = 0 \,,
\label{eq2:Jtests}
\end{align}
reflecting the conservation of the three components of the total angular momentum.   
With the orbits and basis defined according to Fig.\ \ref{fig:orbits}, it is straightforward to show that
\begin{eqnarray}
J &=& J_b \cos \iota +  J_3 \cos \iota_3 \,,
\nonumber \\
0 &=& J_b \sin \iota -  J_3 \sin \iota_3 \,,
\nonumber \\
\Omega_3 &=& \Omega + \pi \,,
\label{eq2:Jtests2}
\end{eqnarray}
where $J_b =  m \eta \sqrt{Gmp}$ and $J_3 = M\eta_3 \sqrt{GMP}$, and that these are valid for all time.  Defining
\begin{align}
\beta &\equiv \frac{J_b}{J_3} = \frac{\sin \iota_3}{\sin \iota} \,,
\nonumber \\
z &\equiv \iota + \iota_3 \,,
\label{eq2:betaz}
\end{align}
we obtain the relations
\begin{equation}
\cot \iota = \frac{\beta + \cos z}{\sin z} \,,  \qquad \cot \iota_3 = \frac{\beta^{-1} + \cos z}{\sin z} \,,
\label{eq2:inclinations}
\end{equation}
so that only the {\em relative} inclination $z$ between the two orbits is dynamically relevant; given an evolution for $z$ and $\beta$, the individual orbital inclinations can be recovered algebraically from Eqs.\ (\ref{eq2:inclinations}).

The equations of motion also admit a conserved total energy, in other words
\begin{equation}
\frac{d}{dt} \left ( E_{K,{\rm in}} + E_{K,{\rm out}} + E_{\rm Int} \right ) =0 \,,
\label{eq:Econserved}
\end{equation}
where 
\begin{align}
 E_{K,{\rm in}}&= \frac{1}{2} \eta m v^2 - \frac{G\eta m^2}{r} \,,
 \nonumber \\
 E_{K,{\rm out}} &= \frac{1}{2} \eta_3M V^2 - \frac{G\eta_3 M^2}{R} \,,
 \nonumber \\
 E_{\rm Int} & = \eta \eta_3 \frac{GM^2}{R}  \sum_{\ell = 1}^\infty \frac{(2\ell +1)!!}{(\ell+1) !} \left ( \frac{r}{R} \right )^{\ell+1} g_\ell
 \nonumber \\
& \qquad 
\times n^{L+1} N^{\langle (L+1) \rangle} \,.
\label{eq:Econserved2}
\end{align}
When we substitute the definitions of the orbit elements into these expressions for the energy, we obtain
\begin{align}
 E'_{K,{\rm in}}&=- \frac{G\eta m^2}{2a} \,,
 \nonumber \\
 E'_{K,{\rm out}}&= - \frac{G\eta_3 M^2}{2A} \,,
 \nonumber \\
 E'_{\rm Int} & = \eta \eta_3 {GM^2}  \sum_{\ell = 1}^\infty  \frac{a^{\ell+1}}{P^{\ell +2}}  g_\ell P_{\ell +1} (\zeta)
 \nonumber \\
 & \quad \times (1+E\cos F)^{\ell+2}(1-e\cos u)^{\ell+1}
 \,.
\label{eq:Econserved3}
\end{align}
where $P_{\ell +1}$ is the Legendre polynomial, and 
\begin{align}
\zeta &\equiv \bm{n} \cdot \bm{N} 
\nonumber \\
&= - \frac{\cos u -e}{1-e \cos u} \biggl ( \cos \omega \cos (F+\omega_3)
\nonumber \\
& \quad\quad +\cos z \sin \omega \sin (F+\omega_3) \biggr )
\nonumber \\
& \quad 
+ \frac{\sqrt{1-e^2} \sin u}{1-e \cos u} \biggl ( \sin \omega \cos (F+\omega_3)
\nonumber \\
& \quad\quad
-\cos z \cos \omega \sin (F+\omega_3) \biggr )
\,.
\end{align}
The interaction energy in these expressions has a curious property.  The calculation of $dE_{\rm Int}/dt$ in Eq.\ (\ref{eq:Econserved2}) involves simply replacing  $\bm{x}$ with $\bm{v}$ and  $\bm{X}$ with $\bm{V}$ in an appropriate manner.  But the calculation of  $dE'_{\rm Int}/dt$ is more involved; in addition to the explicit derivatives with respect to $u$ and $F$, there are derivatives of the orbit elements, such as $dp/dt$ and $d\omega/dt$.  However, after inserting the Lagrange planetary equations for these derivatives, there results a tremendous cancellation, so that the final expression has the form
\begin{equation}
\frac{dE'_{\rm Int}}{dt} \equiv \dot{E}'_{\rm Int} = \frac{\partial E'_{\rm Int}}{\partial u} \left (\frac{du}{dt} \right )_{\rm K}  + \frac{\partial E'_{\rm Int}}{\partial F} \left (\frac{dF}{dt} \right )_{\rm K} \,,
\label{eq:dEprimedt}
\end{equation}
where $(du/dt)_K$ is the Keplerian expression of Eq.\ (\ref{eq:dudtK}) and $(dF/dt)_K$ is the outer orbit analogue of $(df/dt)_K$ of Eq.\ (\ref{eq:dfdt}). 
This is as it must be, because the interaction energy depends only on the coordinates $\bm{x}$ and $\bm{X}$, and therefore its time derivative is purely kinematical ($\bm{x} \to \bm{v}$, $\bm{X} \to \bm{V}$), and cannot involve the dynamical equations.  It is simple to show that the result in Eq.\ (\ref{eq:dEprimedt}) agrees with calculating $dE_{\rm Int}/dt$ in Eq.\ (\ref{eq:Econserved2}) and {\em then} substituting the osculating orbit elements.   On the other hand, the time derivatives of the two Keplerian contributions to the energy do involve $da/dt$ and $dA/dt$ and the Lagrange planetary equations, because the original expressions involve $\bm{v}$ and $\bm{V}$.  As a result,  the conservation of total energy must be expressed in the form
\begin{align}
- \frac{d}{dt} \left [ \frac{G\eta m^2}{2a} + \frac{G\eta_2 M^2}{2A} \right ] + \dot{E}'_{\rm Int} =0 \,.
\label{eq:energycons}
\end{align}
Notice that no such complications affect the conservation equations for angular momentum (\ref{eq2:Jtests}) because the interaction terms actually cancel.  The conservation equations (\ref{eq2:Jtests}) and (\ref{eq:energycons}) will be useful checks of our higher-order calculations.

\section{The two-timescale method and orbital averaging}
\label{sec:twotimescale}

\subsection{Brief review of the two-timescale method}
\label{sec:twotimescalereview}

We now wish to obtain the secular evolution of the orbital elements to second order in the orbital perturbations.  This is done using a two-timescale analysis; a brief summary can be found in Appendix A of  \cite{2021PhRvD.103f3003W}; see also \cite{1978amms.book.....B,1990PhRvD..42.1123L,2004PhRvD..69j4021M,2008PhRvD..78f4028H,2017PhRvD..95f4003W}.

Each planetary equation can be written in the generic form
\begin{align}
\frac{dX_\alpha(t)}{dt} &= \varepsilon Q_\alpha (X_\beta(t),t)  \,,
\end{align}
where the $Q_\alpha$ denote the right-hand sides of the Lagrange planetary equations, $\varepsilon$ is a small parameter that characterizes the perturbation.  The solutions will have pieces that vary on a long, secular timescale, of order $1/\varepsilon$ times the orbital timescales, plus periodic pieces that vary on the orbital timescales.  By defining the long-timescale variable $\theta = \varepsilon t$, treating the two variables as independent, and splitting each element into an average part $\tilde{X}_\alpha$ and an ``average-free'' part $Y_\alpha$, 
\begin{equation}
X_\alpha (\theta,t) \equiv \tilde{X}_\alpha (\theta) + \varepsilon Y_\alpha (\tilde{X}_\beta (\theta), t), 
\end{equation}
we can separate each equation into one for the secular evolution of $\tilde{X}_\alpha$ and one for the periodic evolution of $Y_\alpha$, given by
\begin{subequations}
\begin{align}
\frac{d\tilde{X}_\alpha}{d\theta} &= \langle Q_\alpha (\tilde{X}_\beta + \varepsilon Y_\beta, t) \rangle \,,
\label{eq2:aveq}\\
\frac{\partial Y_\alpha}{\partial t} &= {\cal AF} \left (Q_\alpha (\tilde{X}_\beta + \varepsilon Y_\beta, t) \right )  - \varepsilon \frac{\partial Y_\alpha}{\partial \tilde{X}_\gamma} \frac{d\tilde{X}_\gamma}{d\theta} \,,
\label{eq2:avfreeeq}
\end{align}
\label{eq2:maineq}
\end{subequations}
where the average ($\langle \, \rangle$) and average-free ($\cal{AF}$) parts of a function $A$ are defined by
\begin{align}
\langle A \rangle &\equiv \frac{1}{T} \int_0^{T} A(\theta,t) dt \,,
\nonumber \\
 {\cal AF}(A) &\equiv  A(\theta,t) - \langle A \rangle  \,,
\end{align}
where $T$ is a suitable number of periods related to the short-timescale variable $t$, yet short compared to the perturbation timescale $t/\varepsilon$.  Equation (\ref{eq2:avfreeeq}) can be integrated with respect to $t$ (holding $\theta$ fixed), with the constant of integration determined by requiring that $Y_\alpha$ be average-free.
Taylor expanding Eq.\ (\ref{eq2:avfreeeq}) with respect to $\varepsilon$, we obtain
\begin{equation}
Y_\alpha^{(0)} (t) =  {\cal AF} \int_0^t  {\cal AF} \left (Q^{(0)}_\alpha (t') \right ) dt' + O(\varepsilon) \,,
\label{eq:Ydef}
\end{equation}
where $Q^{(0)}_\alpha \equiv Q_\alpha (\tilde{X}_\beta, t)$.
Taylor expanding Eq.\ (\ref{eq2:aveq}) and substituting for $Y_\alpha^{(0)}$, we obtain, to second order in $\varepsilon$,
\begin{eqnarray}
\frac{d\tilde{X}_\alpha}{dt} &=& \varepsilon \left\langle Q_\alpha^{(0)} + \varepsilon Q_{\alpha , \beta} ^{(0)} Y_\beta^{(0)} \right\rangle 
\nonumber \\
&=&
\varepsilon \left\langle Q_\alpha^{(0)} \right\rangle 
 + \varepsilon^2 \left\langle {\cal AF} \left (Q_{\alpha,\beta}^{(0)} \right) \int_0^t {\cal AF}\left (Q_{\beta}^{(0)} \right )dt' \right\rangle 
 \nonumber
 \\
&& \quad
+O(\varepsilon^3)\,,
\label{eq2:dXdtfinal}
\end{eqnarray}
where the subscript $,\beta$ denotes $\partial/\partial \tilde{X}_\beta$  (we sum over repeated indices), and where we have converted from $\theta$ back to $t$.   We note the useful identities 
\begin{align}
\left \langle A \int_0^t B dt' \right \rangle &= - \left \langle B \int_0^t A dt' \right \rangle + T\langle A \rangle\langle B \rangle \,,
\nonumber \\
\left \langle A \times {\cal AF} \left (B \right ) \right \rangle 
&=  \left \langle  {\cal AF} \left (A \right ) \times  {\cal AF} \left ( B \right )  \right \rangle \,.
\label{eq:identities}
\end{align}
We call the second-order corrections of order $\varepsilon^2$ in Eq.\ (\ref{eq2:dXdtfinal}) ``feedback'' effects, because they result from feeding the periodic perturbations of the elements back into the Lagrange planetary equations, and averaging again.  As the summation includes all the orbit elements, there will be feedback effects from both the inner and the outer elements.

\subsection{Averages of functions of orbit elements}
\label{sec:averages}

The foregoing discussion refers to averaging the time derivatives of the orbit elements.  But we
will also need to calculate averages of functions of orbit elements, such as $1/a$ or $\sqrt{Gmp}$, as well as averages of their time derivatives.  For example, in higher-order perturbation theory, it turns out that $\langle 1/a \rangle$ is not exactly the same as $1/\langle a \rangle$.  Here we work out the properties of functions of orbit elements, of which we typically encounter two types, functions of the orbit elements alone,  $F(X_\alpha(t))$, and functions that have explicit time dependence, $G(X_\alpha (t), t)$.  We substitute $X_\alpha (t) = \tilde{X}_\alpha (\theta)+\varepsilon Y^{(0)}_\alpha (\tilde{X}_\beta, t) + \varepsilon^2 Y^{(1)}_\alpha (\tilde{X}_\beta, t) + \dots$ into $F(X_\alpha(t))$, expand in powers of $\varepsilon$, and then average, taking into account that $\langle Y^{(n)}_\alpha \rangle =0$. The result is
\begin{align}
\langle F(X_\alpha(t))\rangle &= \left \langle F(\tilde{X}_\alpha) + \varepsilon F(\tilde{X}_\alpha)_{,\beta} Y^{(0)}_\beta
\right .
\nonumber \\
&\qquad \left . 
+\varepsilon^2 F(\tilde{X}_\alpha)_{,\beta} Y^{(1)}_\beta
\right .
\nonumber \\
&\qquad \left . 
+ \frac{\varepsilon^2}{2}  F(\tilde{X}_\alpha)_{,\beta\gamma} Y^{(0)}_\beta  Y^{(0)}_\gamma  + \dots \right \rangle
\nonumber \\
&= F(\tilde{X}_\alpha) + \frac{1}{2} \varepsilon^2 F(\tilde{X}_\alpha)_{,\beta\gamma}  \langle Y^{(0)}_\beta  Y^{(0)}_\gamma \rangle \,,
\label{eq:avfunction}
\end{align}
in other words, through order $\varepsilon$, the average of a function of elements is the function of the averages, but at $O(\varepsilon^2)$ there are corrections.
Similarly, we can evaluate the average of the  time derivative of a function, by writing 
\begin{equation}
\frac{d}{dt} F(X_\alpha(t)) = \varepsilon F(X_\alpha(t))_{,\beta} Q_\beta (X_\gamma(t),t) \,.
\end{equation}
Applying the same substitutions, Taylor expanding both functions, averaging, and making use of the identities (\ref{eq:identities}), we find 
\begin{align}
\left \langle \frac{d}{dt} F(X_\alpha(t)) \right \rangle &=  \frac{d}{dt} F(\tilde{X}_\alpha) 
\nonumber \\
& \quad 
+ \frac{1}{2} \epsilon^2 \frac{d}{dt} \left [ F(\tilde{X}_\alpha)_{,\beta\gamma}  \langle Y^{(0)}_\beta  Y^{(0)}_\gamma \rangle \right ] \,,
\end{align}
in agreement with Eq.\ (\ref{eq:avfunction}).
Thus, through second order in perturbation theory, the average of the time derivative of a function of orbit elements is the time derivative of the same function of the averages.
However, no such simplifications occur when averaging a function $G(X_\alpha(t),t) $ with explicit time dependence.  Thus for example, the average of $\dot{E}'_{\rm Int}$ from the energy conservation expression (\ref{eq:dEprimedt}) yields
\begin{align}
\langle \dot{E}'_{\rm Int} (X_\alpha(t),t) \rangle &= \langle  \dot{E}'_{\rm Int} (\tilde{X}_\alpha,t) \rangle + \varepsilon \langle  \dot{E}'_{\rm Int} (\tilde{X}_\alpha,t)_{,\beta} Y^{(0)}_\beta \rangle
\nonumber \\
& \qquad 
 + O(\varepsilon^2) \,.
 \label{eq:avEdot}
\end{align}
The first term might vanish if $\dot{E}'_{\rm Int} (\tilde{X}_\alpha,t) $ is purely periodic over the averaging time, but otherwise nothing special happens.  As we will see later, there will be non-vanishing contributions to the averages in Eq.\ (\ref{eq:avEdot}) at order $\varepsilon$ that will exactly cancel second-order contributions from the derivative of the Keplerian energies involving $da/dt$ and $dA/dt$, thus maintaining total
energy conservation.

\subsection{Orbital averaging at second order}
\label{sec:secularapprox}

In order to apply the two-timescale method to hierarchical triples, we must deal with the fact that the short timescale is actually two orbital timescales, with one long compared to the other, though still short compared to the perturbation timescale.  This is handled by a method that is frequently called the ``secular approximation''.    The functions to be averaged consist of a sum of products of functions of the form $A(t) \times M(t)$ where $A(t)$ is periodic on the inner orbital timescale and $M(t)$ is periodic on the outer orbital timescale.  One then  takes the integral over the time $T$, where $T$ is assumed to be some number of outer orbital periods $P_2$, and divides it into a sequence of subintervals of size $P_1$, the inner orbital period. Within each subinterval $q$, one expands the slowly varying function $M(t) = M(t_q) + (t-t_q) \dot{M}_q + \dots$, integrates over the subinterval, and then sums over all the subintervals, converting sums of $M_q$ back into integrals of $M(t)$ over the time $T$ (see Appendix B of \cite{2021PhRvD.103f3003W} for details).  The result is that the average of the product of functions is the product of their individual averages; equivalently one can integrate separately over the inner and outer time variables holding the other time fixed:
\begin{align}
\langle A(t) M(t) \rangle &\equiv \frac{1}{T} \int_0^T   A M dt 
\nonumber \\
&= \langle A(t)  \rangle \langle M(t)  \rangle 
\nonumber \\
&= \frac{1}{T_1 T_2} \int_0^{T_1}  \int_0^{T_2}  A (t') M(t'') dt' dt'' 
\nonumber \\
& = \frac{1}{T_1 T_2} \int_0^{2\pi}  \int_0^{2\pi}  A(u)M(F) \frac{dt}{du} \frac{dt}{dF} du dF \,,
\label{eq:secapprox1}
\end{align}
where in the last line, we have converted from the two times to the two orbital anomalies.  The product $AM$ stands for $Q_\alpha$, which retains this product form even when expanded to second order; this is because, despite being an integral over time, the function $Y_\alpha^{(0)}$ can be expressed in the product form (see Appendix \ref{app:Y}).  However, the functions $dt/du$ and $dt/dF$, given by Eqs.\ (\ref{eq:dfdt}) and (\ref{eq:dudt1}), also depend on the orbit elements, and must also be Taylor expanded about the averaged elements.  The result is
\begin{align}
\langle Q_\alpha \rangle &=  \frac{1}{T_1 T_2} \int_0^{2\pi}  \int_0^{2\pi}  
\left ( Q_\alpha^{(0)} + \varepsilon Q_{\alpha , \beta} ^{(0)} Y_\beta^{(0)} \right )
\nonumber \\
& \quad \times 
\left [ \frac{dt}{du}^{(0)} + \varepsilon \left ( \frac{dt}{du} \right )^{(0)}_{,\beta} Y_\beta^{(0)} \right ]
\nonumber \\
& \quad \times 
\left [ \frac{dt}{dF}^{(0)} +\varepsilon \left ( \frac{dt}{dF} \right )^{(0)}_{,\beta} Y_\beta^{(0)} \right ] du dF
\nonumber \\
&= \frac{1}{T_1 T_2} \int_0^{2\pi}  \int_0^{2\pi}  \left ( \widehat{Q}_\alpha^{(0)} + \varepsilon \widehat{Q}_{\alpha , \beta} ^{(0)} Y_\beta^{(0)} \right ) du dF \,,
\label{eq:avQ1}
\end{align} 
where 
\begin{equation}
 \widehat{Q}_\alpha^{(0)} \equiv  {Q}_\alpha^{(0)} \frac{dt}{du}^{(0)} \frac{dt}{dF}^{(0)} \,,
\end{equation}
and where $dt^{(0)}/du$ and $dt^{(0)}/dF$ are obtained from Eqs.\ (\ref{eq:dfdt}) and (\ref{eq:dudt1}) using averaged orbit elements, 
\begin{align}
\frac{dt}{du}^{(0)} &=\left (\frac{dt}{du} \right )^{(0)}_K \left ( 1 + \varepsilon Z_u \right ) \,,
\nonumber \\
\frac{dt}{dF}^{(0)} &=\left (\frac{dt}{dF} \right )^{(0)}_K \left ( 1 + \varepsilon Z_F \right ) \,,
\end{align}
where the ``Keplerian'' expressions are given by 
\begin{align}
\left (\frac{dt}{du} \right )^{(0)}_K &= \frac{1}{\tilde{n}} (1- \tilde{e} \cos u ) \,,
\nonumber \\
\left (\frac{dt}{dF} \right )^{(0)}_K &= \frac{(1- \tilde{E}^2)^{3/2}}{\tilde{N} (1+\tilde{E} \cos F)^2
} \,,
\end{align}
with $\tilde{n} = (Gm/\tilde{a}^3)^{1/2}$ and $\tilde{N} = (GM/\tilde{A}^3)^{1/2}$, and where
\begin{align}
Z_u & =  \left (\frac{dt}{du} \right )^{(0)}_K \left ( \frac{(1- \tilde{e} \cos u)}{ \sqrt{1-\tilde{e}^2}} Q_\varpi^{(0)} + \frac{\sin u }{ (1-\tilde{e}^2)} Q_e^{(0)} \right )\,,
\nonumber \\
Z_F & = \left (\frac{dt}{dF} \right )^{(0)}_K Q_{\varpi_3}^{(0)} \,.
\end{align}

\begin{widetext}
However, if $Q_\alpha =1$, its average must be unity.  Setting $Q_\alpha = 1$ in the first of Eqs.\ (\ref{eq:avQ1}), we obtain
\begin{align}
T_1T_2 &=  \int_0^{2\pi}  \int_0^{2\pi} \biggl \{\left (\frac{dt}{du} \right )^{(0)}_K  \left (\frac{dt}{dF} \right )^{(0)}_K \left ( 1 + \varepsilon Z_u + \varepsilon Z_F \right ) 
+ \varepsilon \left [ \left (\frac{dt}{du} \right )^{(0)}_{K,\beta}  \left (\frac{dt}{dF} \right )^{(0)}_K +
\left (\frac{dt}{du} \right )^{(0)}_K  \left (\frac{dt}{dF} \right )^{(0)}_{K,\beta} \right ] Y^{(0)}_\beta \biggr \} du dF
\nonumber \\
& = \frac{2\pi}{\tilde{n}} \frac{2\pi}{\tilde{N}} + \varepsilon T_1T_2 \left ( \langle Z_u \rangle + \langle Z_F \rangle + \frac{1}{\tilde{n}} \langle \sin u Q_e^{(0)} \rangle + \frac{\sqrt{1-\tilde{E}^2}}{\tilde{N}} 
\left \langle \frac{\sin F (2 + \tilde{E} \cos F)}{(1+\tilde{E} \cos F)^2} Q_E^{(0)} \right \rangle \right )
\,,
\label{eq:T1T2}
\end{align}
(see Appendix \ref{app:timeconversion} for detailed calculations).
Using this to eliminate $T_1T_2$ from Eq.\ (\ref{eq:avQ1}), we find three kinds of terms.
The first is the standard first-order term obtained using the secular approximation, holding the orbit elements fixed,
\begin{equation}
\left (\frac{dX_\alpha}{dt} \right )_{0} =\varepsilon \frac{\tilde{n}}{2\pi} \frac{\tilde{N}}{2\pi} \int_0^{2\pi} \int_0^{2\pi} \widehat{Q}^{(0)}_\alpha   du dF \,,
\label{eq:key1}
\end{equation}
where $\widehat{Q}^{(0)}_\alpha$ is constructed from $Q_\alpha$ using the Keplerian expressions $(dt/du)^{(0)}_K$ and $(dt/dF)^{(0)}_K$; we will use $dt/du$ and $dt/dF$ hereafter to denote those Keplerian expressions.  At order $\varepsilon^2$ there are terms arising from the corrections to $dt/du$ and $dt/dF$. We call these the inner and outer ``time conversion'' terms.
\begin{align}
\left (\frac{dX_\alpha}{dt} \right )_{\rm ConvIn} &= \varepsilon^2 \frac{\tilde{n}}{2\pi} \frac{\tilde{N}}{2\pi} \int_0^{2\pi} \int_0^{2\pi} \widehat{Q}^{(0)}_\alpha 
{\cal AF} \left [\frac{dt}{du} \left ( \frac{(1-\tilde{e} \cos u)}{\sqrt{1-\tilde{e}^2}} {Q}^{(0)}_\varpi + \frac{\sin u}{1-\tilde{e}^2} {Q}^{(0)}_e \right ) \right ] du dF - \frac{\varepsilon^2}{\tilde{n}}\langle {Q}^{(0)}_\alpha \rangle \langle \sin u {Q}^{(0)}_e \rangle \,,
\nonumber \\
\left (\frac{dX_\alpha}{dt} \right )_{\rm ConvOut} &= \varepsilon^2  \frac{\tilde{n}}{2\pi} \frac{\tilde{N}}{2\pi} \int_0^{2\pi} \int_0^{2\pi} \widehat{Q}^{(0)}_\alpha 
{\cal AF} \left [ \frac{dt}{dF} {Q}^{(0)}_{\varpi_3} \right ] du dF
- \varepsilon^2  \frac{\sqrt{1-\tilde{E}^2}}{\tilde{N}}  \langle {Q}^{(0)}_\alpha \rangle 
\left \langle \frac{\sin F (2 + \tilde{E} \cos F)}{(1+\tilde{E} \cos F)^2} Q_E^{(0)} \right \rangle \,.
\label{eq:key2}
\end{align}
Because $Y_\beta^{(0)}$ has a dominant and subdominant contribution (see Eq.\ (\ref{eq:Yfinal})), the $O(\varepsilon)$ term in Eq.\ (\ref{eq:avQ1}) contributes a dominant and subdominant ``feedback'' term.  Converting these into explicit integrals over $u$ and $F$, we obtain
\begin{align}
\left (\frac{dX_\alpha}{dt} \right )_{\rm Dom} &= \varepsilon^2 \frac{\bar{N}}{2\pi} \int_0^{2\pi}  \left \{ \left [\frac{\bar{n}}{2\pi} \int_0^{2\pi}  \widehat{Q}_{\alpha , \beta}^{(0)} du -  \frac{dt}{dF} \left \langle {Q}_{\alpha , \beta}^{(0)} \right \rangle   \right ]
\int_0^F \left [\frac{\bar{n}}{2\pi} \int_0^{2\pi}  \widehat{Q}_{ \beta}^{(0)} du - \frac{dt}{dF'} \left \langle {Q}_{ \beta}^{(0)} \right \rangle   \right ] dF' \right \} dF \,,
\nonumber \\
\left (\frac{dX_\alpha}{dt} \right )_{\rm Sub} &=   \varepsilon^2 \frac{\bar{n}}{2\pi}\frac{\bar{N}}{2\pi}
\int_0^{2\pi}  \left \{\int_0^{2\pi} \left [ \widehat{Q}_{\alpha , \beta}^{(0)} -  \frac{\bar{n}}{2\pi}  \frac{dt}{du} \int_0^{2\pi}  \widehat{Q}_{\alpha , \beta}^{(0)} du' \right ] du
\int_0^u \left [ \widehat{Q}_{ \beta}^{(0)}  - \frac{\bar{n}}{2\pi} \frac{dt}{du'}  \int_0^{2\pi}  \widehat{Q}_{\beta}^{(0)} du'' \right ] du' 
\right \}   \frac{dF}{dt}  dF \,,
\label{eq:key3}
\end{align}
where
\begin{align}
 \left \langle {Q}_{\alpha , \beta}^{(0)} \right \rangle & =  \frac{\bar{n}}{2\pi}\frac{\bar{N}}{2\pi}  \int_0^{2\pi}  \int_0^{2\pi}  
 \widehat{Q}_{\alpha , \beta}^{(0)} du dF \,,
 \nonumber \\
\left \langle {Q}_{ \beta}^{(0)} \right \rangle  & =  \frac{\bar{n}}{2\pi}\frac{\bar{N}}{2\pi}  \int_0^{2\pi}  \int_0^{2\pi} 
 \widehat{Q}_{\beta}^{(0)} du dF \,.
\end{align}
Equations (\ref{eq:key1}) - (\ref{eq:key3}) are the key tools for calculating the long-timescale evolution of the orbit elements to second order in perturbation theory.
\end{widetext}

 \begin{figure}[t]
\begin{center}

\includegraphics[width=3.4in]{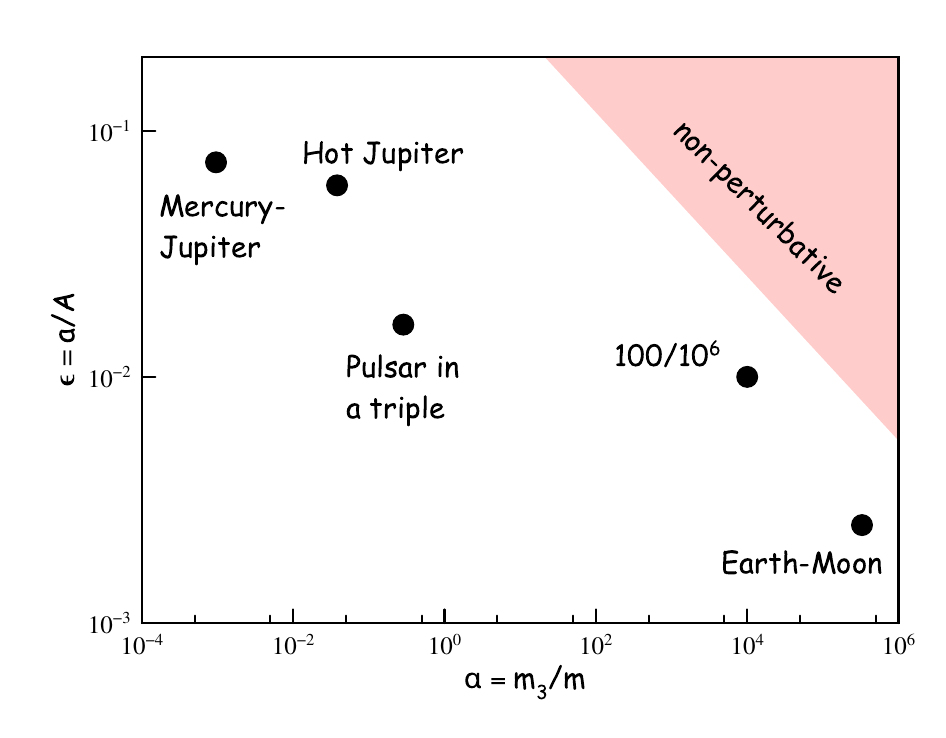}

\caption{Parameter space for the perturbation theory of hierarchical triples.  The shaded region is non-perturbative, with $\alpha \epsilon^3 > 1$.
\label{fig:paramspace} }
\end{center}
\end{figure}

\medskip
\section{Expectations and implementation}

In this section we describe the domain of validity of the perturbative approach and the contributions to be expected at various orders for the secular evolution equations.  The detailed formulae, which are long, will be displayed in Appendix \ref{app:results}.  

In order to see what contributions we expect from first and  second-order perturbations, it is useful to express the Lagrange planetary equations in terms of a dimensionless time $\tau$ given by 
\begin{equation}
\tau \equiv  \frac{t}{P_{\rm in}} = \frac{\tilde{n} t}{2\pi}  \,.
\end{equation}
Then, from Eqs.\ (\ref{eq2:lagrange}), it is straightforward to show that the rates of change of inner and outer orbit elements at the leading, quadrupole order scale according to
\begin{align}
\frac{dX_{\alpha,{\rm in}}}{d\tau} &\sim \alpha \epsilon^3 \,,
\nonumber \\
\frac{dX_{\alpha,{\rm out}}}{d\tau} &\sim \alpha \beta \epsilon^3 \,,
\end{align}
where $\alpha = m_3/m$ and $\epsilon = a/A$. 
For the specific dimensionful orbit elements $p$ and $P$, $X_\alpha$ denotes the logarithm of the element.

There are two fundamental small parameters in the problem.  The first is the amplitude of the quadrupole perturbations, given by $A_Q = \alpha \epsilon^3$.   The second is $\epsilon$ itself, which describes the scaling of the various multipolar contributions, and is small according to the hierarchical assumption.    Notice that the hierarchical assumption also guarantees that the ratio of the inner to the outer orbital period will be small, since
\begin{equation}
\frac{P_{\rm in}}{P_{\rm out}} = \left ( 1+ \frac{m_3}{m} \right )^{1/2} \left ( \frac{a}{A} \right )^{3/2} = \left ( A_Q + \epsilon^3 \right )^{1/2} \,.
\end{equation}

Figure \ref{fig:paramspace} shows the parameter space for the problem, with $\alpha$ plotted vs. $\epsilon$.  The shaded region indicates where $A_Q$ is larger than unity, where a perturbative analysis is not valid.   The parameter $\epsilon$ is chosen to be smaller than $1/5$ in the figure, indicating systems that are reasonably hierarchical.   Figure \ref{fig:paramspace} indicates the location of some representative systems, such as the Sun-Mercury-Jupiter system, the Earth-Moon-Sun system, exoplanet systems denoted ``Hot Jupiters", with a star-Jupiter system perturbed by a distant low-mass star \cite{2011Natur.473..187N}, the pulsar in a triple system J0337+1715 \cite{2014Natur.505..520R}, and a $100 \, M_\odot$ compact binary system orbiting a million solar mass black hole (denoted $100/10^6$).   

Another important parameter is $\beta$, the ratio between the outer and the inner orbital angular momenta (Eq.\ (\ref{eq2:betaz})).  In terms of $\alpha$ and $\epsilon$, $\beta$ is given by
\begin{equation}
\beta = \eta \frac{\sqrt{1+\alpha}}{\alpha} \epsilon^{1/2} \left ( \frac{1-e^2}{1-E^2} \right )^{1/2} \,.
\end{equation}
It is not a perturbative parameter, and indeed can be large or small depending on the system.  

At first order in perturbation theory, for the inner orbit elements, we expect to obtain the conventional quadrupole ($\alpha \epsilon^3$),  octupole  ($\alpha \epsilon^4$), hexadecapole  ($\alpha \epsilon^5$), and dotriacontapole ($\alpha \epsilon^6$) contributions.  For the outer orbit elements, the amplitudes are multiplied by $\beta$.  
Tables \ref{tab:table1} and \ref{tab:table2} show the amplitudes expected at first order for the various multipoles. 

One of the motivations for Paper I was the recognition that dotriacontapole contributions at $\alpha \epsilon^6$ would have the same $\epsilon^6$ scaling as contributions at second order in the quadrupole terms (``quadrupole-squared'' effects), or $(\alpha \epsilon^3 )^2$.   However, we found that, when incorporating the ``feedback'' effect of periodic perturbations of the inner orbit elements at second order, the double orbit average produced terms larger than expected by a factor of $P_{\rm out}/P_{\rm in} = \tilde{n}/\tilde{N} \sim (1+\alpha)^{-1/2} \epsilon^{-3/2}$, resulting in an effective amplitude of  $(\alpha \epsilon^3) \times (\alpha \epsilon^3) \times (1+\alpha)^{-1/2} \epsilon^{-3/2} \sim \alpha^2 (1+\alpha)^{-1/2} \epsilon^{9/2}$.  We call these the ``dominant inner element feedback'' terms, as calculated in Paper I.  The ``subdominant inner element feedback'' terms will have the expected amplitudes of $\alpha^2 \epsilon^6$.  Are there additional contributions at second order that are quadratic in $\alpha$ with a power of $\epsilon$ smaller than six?  In fact there are several.

We first focus on the evolution of inner orbit elements.
The dominant terms from the ``outer element feedback'' will have the form $(\alpha \epsilon^3) \times (\alpha \beta \epsilon^3) \times (1+\alpha)^{-1/2} \epsilon^{-3/2} \sim \alpha \eta \epsilon^{5}$, which is comparable to hexadecapole terms at first order.  On the other hand, the ``subdominant outer element feedback'' terms will be of order  $(\alpha \epsilon^3) \times (\alpha \beta \epsilon^3) \sim \alpha \eta \epsilon^{13/2}$, which is beyond the (arbitrary) cut-off that we have chosen for this paper.   However, additional contributions at a relevant order come from quadrupole-octopole ``cross-feedback'' effects, including feedback of periodic octupole perturbations into the quadrupole terms, and feedback of periodic quadrupole perturbations into the octupole terms.  For dominant inner element feedback, these terms will be of order
$(\alpha \epsilon^3) \times (\alpha  \Delta \epsilon^4) \times  (1+\alpha)^{-1/2} \epsilon^{-3/2} \sim \alpha^2 (1+\alpha)^{-1/2} \Delta \epsilon^{11/2}$, 
while for dominant outer element feedback, they will be of order $(\alpha \epsilon^3) \times (\alpha  \Delta \beta \epsilon^4) \times  (1+\alpha)^{-1/2} \epsilon^{-3/2} \sim \alpha \eta \Delta \epsilon^{6}$.
All subdominant contributions of this type will be below the level of $\epsilon^6$ and will not be computed.

Additional contributions at orders $\epsilon^6$ and above come from corrections to the conversion from time to the anomalies $u$ and $F$, which are of the form
$(dt/du)_K Q_\varpi \sim (dt/du)_K Q_e \sim  \alpha \epsilon^3$ for the inner eccentric anomaly $u$ and 
$(dt/dF)_K Q_{\varpi_3} \sim (\tilde{n}/\tilde{N}) \alpha \beta \epsilon^3 \sim \eta \epsilon^2$ for the outer true anomaly.   As a result, the time conversion corrections applied to quadrupole perturbations lead to terms with amplitudes $\alpha \epsilon^3 \times \eta \epsilon^2 \sim \alpha \eta \epsilon^5$ and 
$\alpha \epsilon^3 \times \alpha \epsilon^3 \sim \alpha^2 \epsilon^6$, respectively.  In addition, the outer true anomaly {\em octupole} correction $\sim \eta \Delta \epsilon^3$ applied to quadrupole perturbations (and vice versa) leads to terms with amplitudes $\alpha \epsilon^3 \times  \eta \Delta \epsilon^3 \sim \alpha \eta \Delta \epsilon^6$.  All other contributions are of order $\epsilon^{13/2}$ or higher, and will not be computed.  Table \ref{tab:table1} summarizes these contributions.

\begin{widetext}
\begin{center}
\begin{table}[t]
\caption{Inner orbit elements}
\medskip
\begin{tabular}{l@{\hskip 1 cm}l@{\hskip 0.5cm}c@{\hskip 0.5cm}l@{\hskip 0.5cm}c}
\hline
\vspace{-0.3cm}\\
Contribution&Nominal&Dominant-&Final&\\
&Amplitude&Subdominant&Amplitude&$da/d\tau$\\
\vspace{-0.3cm}\\
\hline  
\vspace{-0.3cm}\\
\multicolumn{3}{l}{\em First-order perturbations}\\
Quadrupole&$\alpha \epsilon^3$&NA&{$\alpha \epsilon^3$}&0\\
Octupole&$\alpha\Delta\epsilon^4$&NA&{$\alpha\Delta\epsilon^4$}&0\\
Hexadecapole&$\alpha(1-3\eta)\epsilon^5$&NA&{$\alpha(1-3\eta)\epsilon^5$}&0\\
Dotriacontapole&$\alpha\Delta(1-2\eta)\epsilon^6$&NA&{$\alpha\Delta(1-2\eta)\epsilon^6$}&0\\
\vspace{-0.3cm}\\
\hline
\vspace{-0.3cm}\\
\multicolumn{3}{l}{\em Second-order feedback}\\
$Q_{{\rm in,in}}^{\rm quad} \int^t Q^{\rm quad}_{\rm in}$&$\alpha \epsilon^3 \times \alpha \epsilon^3$
&Dom&{$\alpha^2 (1+\alpha)^{-1/2} \epsilon^{9/2}$}&$0$\\
$Q_{{\rm in,out}}^{\rm quad} \int^t Q^{\rm quad}_{\rm out}$&$\alpha \epsilon^3 \times \alpha \beta \epsilon^3$
&Dom&{${\alpha}\eta \epsilon^{5}$}&$0$\\
$Q_{{\rm in,in}}^{\rm (quad} \int^t Q^{\rm oct)}_{\rm in}$&$\alpha \epsilon^3 \times \alpha \Delta \epsilon^4$
&Dom&{$\alpha^2 (1+\alpha)^{-1/2}  \Delta \epsilon^{11/2}$}&$0$\\
$Q_{{\rm in,out}}^{\rm (quad} \int^t Q^{\rm oct)}_{\rm out}$&$\alpha \epsilon^3 \times \alpha \Delta \beta \epsilon^4$
&Dom&{$\alpha\eta\Delta\epsilon^{6}$}&$0$\\
$Q_{{\rm in,in}}^{\rm quad} \int^t Q^{\rm quad}_{\rm in}$&$\alpha \epsilon^3 \times \alpha \epsilon^3$&Sub&{$\alpha^2 \epsilon^6$}&$\ne 0$\\

\vspace{-0.3cm}\\
\hline
\vspace{-0.3cm}\\
\multicolumn{3}{l}{\em Second-order time conversion}\\
$Q_{\rm in}^{\rm quad}  (dt/dF)^{\rm quad}$&$\alpha \epsilon^3 \times \eta \epsilon^2$&NA&{$\alpha \eta \epsilon^{5}$}&$\ne 0$\\
$Q_{\rm in}^{\rm quad}  (dt/df)^{\rm quad}$&$\alpha \epsilon^3 \times \alpha \epsilon^3$&NA&{$\alpha^2 \epsilon^6$}&$\ne 0$\\
$Q_{\rm in}^{\rm (quad}  (dt/dF)^{\rm oct)}$&$\alpha \Delta \epsilon^4 \times  \eta \epsilon^2$&NA&{$\alpha \eta \Delta \epsilon^{6}$}&$\ne 0$\\
\vspace{-0.3cm}\\
\hline
\end{tabular}
\label{tab:table1}
\end{table}
\end{center}

\begin{center}
\begin{table}[t]
\caption{Outer orbit elements}
\medskip
\begin{tabular}{l@{\hskip 1 cm}l@{\hskip 0.5cm}c@{\hskip 0.5cm}l@{\hskip 0.5cm}c}
\hline
\vspace{-0.3cm}\\
Contribution&Nominal&Dominant-&Final\\
&Amplitude&Subdominant&Amplitude&$dA/d\tau$\\
\vspace{-0.3cm}\\
\hline  
\vspace{-0.3cm}\\
\multicolumn{3}{l}{\em First-order perturbations}\\
Quadrupole&$\alpha \beta \epsilon^3$&NA&{$\alpha \beta  \epsilon^3$}&$0$\\
Octupole&$\alpha \Delta \beta \epsilon^4$&NA&{$\alpha\Delta \beta \epsilon^4$}&$0$\\
Hexadecapole&$\alpha(1-3\eta)\beta \epsilon^5$&NA&{$\alpha(1-3\eta) \beta \epsilon^5$}&$0$\\
Dotriacontapole&$\alpha\Delta(1-2\eta) \beta \epsilon^6$&NA&{$\alpha\Delta(1-2\eta) \beta \epsilon^6$}&$0$\\
\vspace{-0.3cm}\\
\hline
\vspace{-0.3cm}\\
\multicolumn{3}{l}{\em Second-order feedback}\\
$Q_{{\rm out,in}}^{\rm quad} \int^t Q^{\rm quad}_{\rm in}$&$\alpha \beta \epsilon^3 \times \alpha \epsilon^3$
&Dom&{$\alpha^2 (1+\alpha)^{-1/2} \beta \epsilon^{9/2}$}&$0$\\
$Q_{{\rm out,out}}^{\rm quad} \int^t Q^{\rm quad}_{\rm out}$&$\alpha\beta  \epsilon^3 \times \alpha \beta \epsilon^3$
&Dom&{${\alpha}\eta \beta  \epsilon^{5}$}&$\ne 0$\\
$Q_{{\rm out,in}}^{\rm (quad} \int^t Q^{\rm oct)}_{\rm in}$&$\alpha\beta  \epsilon^3 \times \alpha \Delta \epsilon^4$
&Dom&{$\alpha^2 (1+\alpha)^{-1/2} \Delta \beta \epsilon^{11/2}$}&$0$\\
$Q_{{\rm out,out}}^{\rm (quad} \int^t Q^{\rm oct)}_{\rm out}$&$\alpha \beta  \epsilon^3 \times \alpha \Delta \beta \epsilon^4$
&Dom&{$\alpha\eta\Delta \beta \epsilon^{6}$}&$\ne 0$\\
$Q_{{\rm out,in}}^{\rm quad} \int^t Q^{\rm quad}_{\rm in}$&$\alpha \beta \epsilon^3 \times \alpha \epsilon^3$&Sub&{$\alpha^2  \beta \epsilon^6$}&$\ne 0$\\
\vspace{-0.3cm}\\
\hline
\vspace{-0.3cm}\\
\multicolumn{3}{l}{\em Second-order time conversion}\\
$Q_{\rm out}^{\rm quad}  (dt/dF)^{\rm quad}$&$\alpha \beta \epsilon^3 \times \eta \epsilon^2$&NA&{$\alpha \eta \beta \epsilon^{5}$}&$\ne 0$\\
$Q_{\rm out}^{\rm quad}  (dt/df)^{\rm quad}$&$\alpha\beta  \epsilon^3 \times \alpha \epsilon^3$&NA&{$\alpha^2 \beta \epsilon^6$}&$\ne 0$\\
$Q_{\rm out}^{\rm (quad}  (dt/dF)^{\rm oct)}$&$\alpha \Delta \beta \epsilon^4 \times  \eta \epsilon^2$&NA&{$\alpha \eta \Delta \beta \epsilon^{6}$}&$\ne 0$\\
\vspace{-0.3cm}\\
\hline
\end{tabular}
\label{tab:table2}
\end{table}
\end{center}
\end{widetext}

For the evolution of the outer orbit elements, the arguments are the same, with the only difference being that the evolutions (in terms of the dimensionless time $\tau$) are multiplied by $\beta$ (see Table \ref{tab:table2}).

The detailed evolutions calculated using our two-timescale method agree with these qualitative arguments.  As the expresions are quite long, we have displayed most of them in Appendix \ref{app:results}.   The quadrupole-octupole cross terms are exceedingly long, so we have not displayed them, but instead have posted all the results 
\href{https://github.com/landenconway/Three-Body-Secular-Equations}{here}.

The two pericenter elements $\omega$ and $\omega_3$ require special treatment, in that they also involve the evolution of the nodal angle $\Omega$ (recall that $d\Omega/d\tau = d\Omega_3/d\tau$).  It turns out that they take a relatively simple form if we first extract suitable expressions involving $d\Omega/d\tau$, and display the remainders.   

In carrying out the explicit integrations in Eqs.\ (\ref{eq:key1}) - (\ref{eq:key3}), we found it useful to convert the inner orbit variables from the true anomaly $f$ to the eccentric anomaly $u$, using the relations (\ref{eq:radiusu}) -- (\ref{eq:dudtK}).
Because the multipole expansion involves positive powers of $r$, this conversion avoids the presence of negative powers of $(1+e \cos f)$ which complicates and slows the computation of integrals using algebraic software such as Mathematica and Maple.  Since the outer radius $R$ appears in the expansions with negative powers, we stick with the true anomaly $F$ for the outer orbit.  

\section{Discussion}

We have obtained a perturbative solution for the long-term evolution of the averaged orbit elements of hierarchical triple systems that is complete through second-order in the perturbation parameter $A_Q = \alpha \epsilon^3$ and through order $\epsilon^6$ in the multipolar expansion parameter.  Apart from requiring that $A_Q \ll 1$ and $\epsilon \ll 1$, restricting the domain of validity of our method to the regions shown in Fig.\ \ref{fig:paramspace}, we place no restrictions on the masses of the bodies or on the inclinations or orientations of the orbits.  The outer body can be a low-mass perturber or it can be a supermassive black hole that dominates the dynamics.   We obtain evolutions for both the inner binary and the outer body.

Formally, we place no restrictions on the orbital eccentricities $e$ and $E$; in practice, however we must acknowledge an important caveat regarding $E$.  The secular approximation requires that the perturbation caused by the third body vary slowly over an orbit of the inner binary so that a low-order Taylor expansion of the perturbing functions is a decent approximation (see Sec.\ \ref{sec:secularapprox}).  But for any finite $E$, the outer perturbation consists of a sequence of harmonics in time, with periods $\sim  P_{\rm out}/s$, and amplitudes $\sim E^s$, where $s$ is an integer.   The secular approximation will begin to fail (and the possibility of resonant interaction with the inner binary will begin to emerge) when $P_{\rm out}/s \sim P_{\rm in}$, i.e. when $s \sim \epsilon^{-3/2}$.  For $E \sim 0.6$ and $\epsilon \sim 1/10$, this implies a perturbation from that harmonic with an amplitude $\sim E^s \sim 10^{-7}$ times the quadrupole amplitude, or $10^{-4}$ times the dotriacontapole amplitude.   For smaller values of $\epsilon$ the effect is even smaller.   Of course, orbital resonance effects can be physically significant even for the smallest perturbation, and thus users of our results should be cognizant of this caveat.

In addition to the standard first-order sequence of multipolar perturbations with amplitudes $\propto \epsilon^q$, with $q = 3,\, 4,\, 5$ and $6$ through dotriacontapole order, we find additional contributions at second-order with $q = 5$ and $6$, as well as contributions at odd-half orders, $q = 9/2$ and $11/2$.   All other contributions at second order will have $q = 13/2$ and higher.  

There is however, a higher-order contribution that will show up at $q = 6$.  This comes from third-order perturbation theory, involving quadrupole cubed terms.  Nominally, these would contribute at order $\alpha^3 \epsilon^9$, but because they involve averaging two integrals over time, such as $\langle Q_{\alpha,\beta\gamma} \int Q_\beta \int Q_\gamma \rangle$ or $\langle Q_{\beta,\gamma} \int Q_{\alpha,\beta} \int Q_\gamma \rangle$, there is a double boost in amplitude by the factor $(P_{\rm out}/P_{\rm in})^2$, so that there is a doubly dominant contribution with amplitude $[(1+\alpha)^{-1/2} \epsilon^{-3/2} ]^2 \alpha^3 \epsilon^9 \sim \alpha^3 (1+\alpha)^{-1} \epsilon^6$.   Other third-order contributions will have $q > 6$.  Evaluating these very complicated contributions will be left to future work.  At third order, one may also have to worry about corrections to averages of functions of orbit elements.

An unexpected result of our calculations is the presence of long-term variations in the averaged semimajor axes $a$ and $A$.  These variations do not occur at first order in perturbation theory: in Appendix \ref{app:semimajor} we prove this to arbitrary multipole orders.  They are also constant in the dominant $Q^2$ terms with inner element feedback [see Eqs.\ (\ref{eq:QQtermsDom})].  Long term variations in $a$ and $A$ first show up in the $Q^2$ terms with outer element feedback and outer time conversion, Eqs.\ (\ref{eq:QQtermsHex}), with amplitude $\propto \alpha \eta \epsilon^5$, the same level as the first-order hexadecapole terms (see Eq.\ ( 2.28) of \cite{2017PhRvD..96b3017W}).   The terms vanish in the extreme-mass ratio limit for the inner binary, $\eta \to 0$.   Variations in $a$ and $A$ also show up in sub-dominant $Q^2$ terms and in dominant quadrupole-octopole cross-terms with outer element feedback.  In all cases, we have verified that total energy conservation, as embodied in Eq.\ (\ref{eq:energycons}), is satisfied. 
As discussed in the Introduction, this is not in direct conflict with historical theorems on secular effects on the semimajor axes of hierarchical triples.  We will address the astrophysical implications of these effects in future publications.
An interesting question is how these second-order perturbative effects can be treated in the conventional Hamiltonian framework.

\acknowledgments

This work was supported in part by the National Science Foundation,
Grants No.\ PHY 19-09247 and PHY 22-07681.   We are grateful for the hospitality of  the Institut d'Astrophysique de Paris where part of this work was carried out.


\appendix

\begin{widetext}

\section{Summary of the results and a handbook for their use}
\label{app:results}

In this appendix we display the evolution equations for the averaged orbit elements.  All new contributions through second order in perturbation theory, i.e. $O([\alpha \epsilon^3]^2)$ and through effective dotriacontapole order, i.e. $O(\epsilon^6)$ have been obtained, where $\alpha = m_3/m$, and $\epsilon = a/A$.  We also have that $\eta = m_1 m_2/m^2$,
$m = m_1 + m_2$, and $\Delta = (m_2 - m_1)/m$.  The dimensionless parameter $\tau$ is time in units of the inner orbit mean period, i.e.\ $\tau = (t/2\pi)(Gm/\tilde{a}^3)^{1/2}$.  

From a dynamical point of view, only the angle $z$ between the two orbital planes is relevant; the individual angles $\iota$ and $\iota_3$ serve only to orient the triple system relative to the reference coordinate basis, whose $Z$ axis is parallel to the total angular momentum of the system.  These are related by the equations
\begin{align}
\iota + \iota_3 & = z \,,
\nonumber \\
\frac{\sin \iota_3}{\sin \iota} &= \frac{J_b}{J_3} = \beta \,,
\end{align}
where $\beta$ is given by 
\begin{equation}
\beta = \frac{\eta m}{m_3} \left ( \frac{M}{m} \right )^{1/2} \left ( \frac{p}{P} \right )^{1/2}  = \eta \epsilon^{1/2}  \frac{ (1+\alpha)^{1/2} }{\alpha} \left ( \frac{1-e^2}{1-E^2} \right )^{1/2} \,.
\end{equation}
 Recall that $\beta$ may be less than or greater than unity.  All of the evolution equations depend only on the relative inclination angle $z$, except for the equations for $d\Omega/d\tau$ and $d\Omega_3/d\tau$, which have a factor of $\sin \iota$ and $\sin \iota_3$ in the denominator, respectively.  If necessary, each of these factors can be expressed in terms of $z$ and $\beta$ using Eqs.\ (\ref{eq2:inclinations}).  

For each type of perturbative term, we display the results explicitly for only seven of the 10 orbit elements of the two orbits, specifically $a$, $A$, $e$, $\iota$, $\Omega$, $\omega$, and $\omega_3$.  The remaining three equations can be constructed from the identities that arise from the conservation of total angular momentum, Eqs.\ (\ref{eq2:Jtests}).  We have calculated all 10 orbit element evolutions and verified explicitly that they satisfy Eqs.\ (\ref{eq2:Jtests}).  When the elements are the orbit averaged elements, these identities are valid through second order in perturbation theory (see Sec.\ \ref{sec:averages} for discussion):
\begin{align}
\frac{d\Omega_3}{d\tau} & = \frac{d\Omega}{d\tau}  \,,
\nonumber \\
\frac{E}{1-E^2} \frac{dE}{d\tau} &= \frac{1}{2A} \frac{dA}{d\tau} -\beta \left (\sin z \frac{d\iota}{d\tau} -\frac{\cos z}{2a} \frac{da}{d\tau} + \frac{e \cos z}{1-e^2}\frac{de}{d\tau} \right )\,,
\nonumber \\
\frac{d\iota_3}{d\tau} &= \beta \left (\cos z \frac{d\iota}{d\tau} + \frac{\sin z}{2a} \frac{da}{d\tau} - \frac{e \sin z}{1-e^2}\frac{de}{d\tau} \right )\,.
\end{align}

The evolution equations for the averaged elements at these high orders are long and complicated.  In an effort to keep the expressions under control, we use the following notation:
\begin{align}
[A^\pm] &\equiv A^+ + A^- \,,
\nonumber
\\
[ (a \mp b) B^\pm ] &\equiv (a-b) B^+ +(a+b) B^- \,,
\nonumber
\\
[ (a \pm b) B^\pm ] &\equiv (a+b) B^+ +(a-b) B^- \,,
\nonumber
\\
[ (a \pm b) c^\mp B^\pm ] &\equiv (a+b) c^- B^+ +(a-b) c^+ B^-  \,,
\end{align}
and so on; each term is a sum of the expression using the upper sign and the same expression using the lower sign.

\subsection{First-order perturbations}

At first order in perturbation theory, we have the conventional multipolar perturbations, obtained by holding the orbit elements fixed on the right-hand-side of the Lagrange planetary equations, while carrying out a double orbit average (the secular approximation), described in Sec.\ \ref{sec:secularapprox}.  At quadrupole, octopole and hexadecapole orders, with conventions and notation identical to those of this paper, the results can be found in Eqs.\ (2.25), (2.27) and (2.28), respectively, of  
\cite{2017PhRvD..96b3017W}.  At dotriacontapole order, the evolution equations are given by
\begin{align}
\frac{da}{d\tau} & =\frac{dA}{d\tau} = 0 \,, 
\nonumber \\
\frac{de}{d\tau} &= -\frac{105 \pi}{65536} \alpha \epsilon^6 \Delta (1-2\eta) \frac{E (1-e^2)^{1/2}}{(1-E^2)^{9/2}}
\biggl [ (4+3E^2) \biggl \{ -4(8+20e^2+5e^4) \left [(1\mp \cos z) {\cal F}^\pm_1 \sin(\omega \pm \omega_3) \right ]
\nonumber \\
&\qquad
+ 126 e^2 (8+3e^2) \sin^2 z \left [{\cal F}^\pm_2 \sin (3\omega \pm \omega_3)  \right ]
-6930 e^4 \sin^4 z \left [ (1 \mp \cos z)  \sin (5\omega \pm \omega_3) \right ] \biggr \}
\nonumber \\
&\qquad
+ 7 E^2 \biggl \{ 2(8+20e^2+5e^4) \sin^2 z \left [ {\cal F}^\pm_2 \sin (\omega \pm 3\omega_3)  \right ]
-3 e^2 (8+3e^2) \left [ (1 \mp \cos z) {\cal F}^\pm_3  \sin (3\omega \pm 3\omega_3) \right ]
\nonumber \\
&\qquad
-495e^4 \sin^2 z \left [ (1 \mp \cos z)^3 \sin (5\omega \pm 3\omega_3) \right ] \biggr \} \biggr ] \,,
\nonumber \\
\frac{d\iota}{d\tau} &=\frac{105 \pi}{65536} \alpha \epsilon^6 \Delta (1-2\eta) \frac{eE \sin z}{ (1-e^2)^{1/2}(1-E^2)^{9/2}} 
\biggl [ (4+3E^2) \biggl \{ -4(8+20e^2+5e^4) \left [  {\cal F}^\pm_1 \sin(\omega \pm \omega_3)
\right ]
\nonumber \\
&\qquad
+ 42 e^2 (8+3e^2) \left [ (1 \pm 3\cos z){\cal F}^\pm_2 \sin (3\omega \pm \omega_3) \right ]
-1386 e^4 \sin^2 z \left [ {\cal F}^\pm_4  \sin (5\omega \pm \omega_3) \right ] \biggr \}
\nonumber \\
&\qquad
+ 7 E^2 \biggl \{ 2(8+20e^2+5e^4)  \left [ (3 \pm \cos z) {\cal F}^\pm_2 \sin (\omega \pm 3\omega_3)  \right ]
-3 e^2 (8+3e^2) \left [  {\cal F}^\pm_3  \sin (3\omega \pm 3\omega_3 )\right ]
\nonumber \\
&\qquad
-99 e^4  \left [  {\cal F}^\pm_5 \sin (5\omega \pm 3\omega_3)  \right ] \biggr \} \biggr ] \,,
\nonumber \\
\frac{d\Omega}{d\tau} &=\frac{105 \pi}{65536} \alpha \epsilon^6 \Delta (1-2\eta) \frac{eE }{ (1-e^2)^{1/2}(1-E^2)^{9/2}} 
\frac{\sin z}{\sin \iota}
\biggl [ (4+3E^2) \biggl \{ 4(8+20e^2+5e^4) \left [  {\cal F}^\pm_6 \cos(\omega \pm \omega_3)
 \right ]
 \nonumber \\
&\qquad
-42 e^2 (8+3e^2)  \left [{\cal F}^\pm_7 \cos (3\omega \pm \omega_3)  \right ]
+1386 e^4 \sin^2 z \left[ {\cal F}^\pm_4 \cos (5\omega \pm \omega_3) \right ] \biggr \}
\nonumber \\
&\qquad
- 7 E^2 \biggl \{ 2(8+20e^2+5e^4)  \left [ {\cal F}^\pm_7 \cos (\omega \pm 3\omega_3) \right ]
+3 e^2 (8+3e^2) \left [  {\cal F}^\pm_8  \cos (3\omega \pm 3\omega_3)  \right ]
\nonumber \\
&\qquad
-99 e^4  \left [ {\cal F}^\pm_5 \cos (5\omega \pm 3\omega_3)\right ] \biggr \} \biggr ] \,,
\nonumber \\
\frac{d\varpi}{d\tau} &=
 \frac{105 \pi}{65536} \alpha \epsilon^6 \Delta (1-2\eta) \frac{E (1-e^2)^{1/2}}{ e(1-E^2)^{9/2}} 
\biggl [ (4+3E^2) \biggl \{ 4(8+60e^2+25e^4) \left [ (1 \mp \cos z) {\cal F}^\pm_1 \cos(\omega \pm \omega_3) \right ]
\nonumber \\
&\qquad 
- 126 e^2 (8+5e^2) \sin^2 z \left [ {\cal F}^\pm_2 \cos (3\omega \pm\omega_3) \right ]
+6930 e^4 \sin^4 z \left [ (1 \mp \cos z)  \cos (5\omega \pm\omega_3) \right ] \biggr \}
\nonumber \\
&\qquad
- 7 E^2 \biggl \{ 2(8+60e^2+25e^4) \sin^2 z \left [ {\cal F}^\pm_2 \cos (\omega \pm3\omega_3)  \right ]
-3 e^2 (8+5e^2) \left [ (1 \mp \cos z) {\cal F}^\pm_3  \cos (3\omega \pm 3\omega_3) \right ]
\nonumber \\
&\qquad
-495e^4 \sin^2 z \left [ (1 \mp \cos z)^3 \cos (5\omega \pm 3\omega_3)  \right ] \biggr \} \biggr ] \,,
\nonumber \\
\frac{d\varpi_3}{d\tau} &=
\frac{105 \pi}{65536} \alpha \epsilon^6  \frac{\beta \Delta (1-2\eta) e }{ E(1-E^2)^{9/2}(1-e^2)^{1/2}} 
\biggl [ (4+41E^2+8E^4)
\biggl \{ 4(8+20e^2+5e^4) \left [ (1 \mp \cos z) {\cal F}^\pm_1 \cos(\omega \pm \omega_3)
 \right ]
\nonumber \\
&\qquad
-42e^2 (8+3e^2) \sin^2 z \left [{\cal F}^\pm_2 \cos (3\omega \pm \omega_3)  \right ]
+1386e^4 \sin^4 z \left [ (1 \mp \cos z)  \cos (5\omega \pm \omega_3) \right ] \biggr \}
\nonumber \\
&\qquad
- 21 E^2 (1+2E^2) \biggl \{ 2(8+20e^2+5e^4) \sin^2 z \left [ {\cal F}^\pm_2 \cos (\omega \pm 3\omega_3)  \right ]
- e^2 (8+3e^2) \left [ (1 \mp \cos z) {\cal F}^\pm_3  \cos (3\omega \pm 3\omega_3) \right ]
\nonumber \\
&\qquad
-99e^4 \sin^2 z \left [ (1 \mp \cos z)^3 \cos (5\omega \pm 3\omega_3)  \right ] \biggr \} \biggr ] \,,
\label{eq:finallinear}
\end{align}
where
\begin{align}
{\cal F}^\pm_1 &= 1 \mp 28 \cos z - 42 \cos^2 z \pm 84 \cos^3 z + 105 \cos^4 z\,,
\nonumber \\
{\cal F}^\pm_2 &=(1 \mp \cos z)(1 \mp 6 \cos z - 15 \cos^2 z ) \,,
\nonumber \\
{\cal F}^\pm_3 &= (1 \mp \cos z)^2 (1 \pm 3\cos z )(13 \pm 15 \cos z) \,,
\nonumber \\
{\cal F}^\pm_4 &=(1 \mp \cos z)(1 \pm 5 \cos z) \,,
\nonumber \\
{\cal F}^\pm_5 &=(1 \mp \cos z)^3 (3 \pm 5 \cos z) \,,
\nonumber \\
{\cal F}^\pm_6 &=29 \pm 28 \cos z - 378 \cos^2 z \mp 84\cos^3 z +525 \cos^4 z \,,
\nonumber \\
{\cal F}^\pm_7 &=(1 \mp \cos z)( 7 \pm 27 \cos z - 39 \cos^2 z \mp 75 \cos^3 z )\,,
\nonumber \\
{\cal F}^\pm_8 &= (1 \mp \cos z)^2 (5 \mp 42 \cos z - 75 \cos^2 z) \,.
\label{eq:Fcoeffs}
\end{align}

Notice that here, and for the lower multipoles at linear order, the semimajor axes $a$ and $A$ are constant. In Appendix \ref{app:semimajor}, we give a proof that they are constant to all multipole orders.  At second order in perturbation theory, this will no longer always be the case.

Here and for the linear results presented in \cite{2017PhRvD..96b3017W}, we  display equations for the auxiliary pericenter variables $d\varpi/d\tau$ and $d\varpi_3/d\tau$, because these are often simpler.  They are related to the true pericenter angle evolution rates $d\omega/d\tau$ and $d\omega_3/d\tau$ by 
\begin{align}
\frac{d\omega}{d\tau} &= \frac{d\varpi}{d\tau} - \frac{\beta + \cos z}{\sin z} \left ( \sin \iota \frac{d\Omega}{d\tau} \right ) \,,
\nonumber \\
\frac{d\omega_3}{d\tau} &= \frac{d\varpi_3}{d\tau} - \frac{1+ \beta \cos z}{\sin z} \left ( \sin \iota \frac{d\Omega}{d\tau} \right ) \,.
\label{eq:Adomdtau}
\end{align}
Note that the factors of $\beta$ will introduce additional factors of  $\epsilon^{1/2}$ into the pericenter expressions at all multipole orders.  These relations hold only for the linear results; for the second-order results we will quote the pericenter evolutions directly.

\subsection{Second-order perturbations: Dominant $Q^2$ terms}
\label{}
  
The dominant $Q^2$ terms from inner element feedback (labeled ``di''), with amplitudes for the inner elements at $O(\alpha^2  \epsilon^{9/2})$,
were already obtained in \cite{2021PhRvD.103f3003W}, but we reproduce them here for completeness. 
\begin{align}
\frac{da^{di}}{d\tau} & =\frac{dA^{di}}{d\tau} = 0 \,, 
\nonumber \\
\frac{de^{di}}{d\tau} & = \frac{15\pi}{32} \frac{\alpha^2 \epsilon^{9/2}}{(1+\alpha)^{1/2}}  \frac{e(1-e^2)}{(1-E^2)^3} \biggl [  3 (3+2E^2) \cos z \sin^2 z \sin 2\omega
+\frac{5}{2} E^2 H(E) \left [ (1 \mp \cos z)^2 (3\cos z \pm 2)\sin(2\omega \pm 2\omega_3)  \right ] \biggr ]\,,
\nonumber \\
\frac{d \iota^{di}}{d\tau} &= - \frac{15\pi}{32} \frac{\alpha^2 \epsilon^{9/2}}{(1+\alpha)^{1/2}}  
\frac{\sin z}{(1-E^2)^3} \biggl [ 3e^2 (3+2E^2) \cos^2 z  \sin 2\omega
\nonumber \\
& \qquad
+\frac{1}{2} E^2 H(E) \biggl \{ 5e^2 \left [(1 \mp \cos z) (2 \pm 3\cos z)\sin(2\omega \pm 2\omega_3) \right ]
-2(2-17e^2) \cos(z) \sin (2\omega_3) \biggr \} \biggr ]\,,
\nonumber \\
\frac{d\Omega^{di}}{d\tau} &= - \frac{3\pi}{64} \frac{\alpha^2 \epsilon^{9/2}}{(1+\alpha)^{1/2}} 
\frac{1}{(1-E^2)^3} \frac{\sin(z)}{\sin(\iota)} \biggl [ (3+2E^2) \biggl ( 2+33e^2-3(2-17e^2)\cos^2 z +15e^2 (1-3\cos^2 z) \cos 2\omega \biggr )
\nonumber \\
& \qquad
-\frac{5}{2} E^2 H(E) \biggl \{ 5e^2 \left [ (1 \mp \cos z) (1 \pm 9\cos z)\cos(2\omega \pm 2\omega_3) \right ]
+2(2-17e^2) (1-3\cos^2 z) \cos 2\omega_3 \biggr \} \biggr ]\,.
\label{eq:QQtermsDom}
\end{align}
The dominant  $Q^2$ terms from outer element feedback and outer time conversion (labeled ``do''), with amplitudes for the inner elements at $O(\alpha^2 \beta \epsilon^{9/2}) \sim O(\alpha \eta \epsilon^5)$, are given by
\begin{align}
\frac{da^{do}}{d\tau} &=\frac{3 \pi a}{128} \frac{\alpha \eta \epsilon^5 (10 +E^2)(1-e^2)^{1/2}}{(1-E^2)^{7/2}}  
  \biggl [ 6 (2+3e^2) \cos z \sin^2 z \sin 2\omega_3 + 5e^2 \left [ (1 \mp \cos z)^2 (2 \pm 3 \cos z) \sin (2\omega \pm 2\omega_3 ) 
  \right ] \biggr ] \,,
  \nonumber \\
  \frac{dA^{do}}{d\tau} &= - \frac{9\pi A}{16384} \frac{\alpha \eta \beta \epsilon^5 }{(1-e^2)^{1/2}(1-E^2)^{9/2}} \biggl [ 2 \sin^2 z \biggl \{ {\cal G}_{1} \sin 2\omega_3 + E^2(4-E^2) (8+72e^2-41e^4) \sin^2 z \sin 4 \omega_3 \biggr \}
\nonumber \\ 
& \qquad
\pm e^2 (64+16E^2-E^4)  \biggl \{4 {\cal G}^\pm_{2} \sin(2\omega \pm 2\omega_3)
+39e^2 \sin^2 z (1 \mp \cos z)^2 \sin(4\omega \pm 2\omega_3) \biggr \}
\nonumber \\ 
& \qquad
\pm e^2 E^2 (4-E^2) \biggl \{ 4(22-9e^2) \sin^2 z (1 \mp \cos z)^2 \sin(2\omega \pm 4\omega_3)
 +13e^2 (1\mp \cos z)^4 \sin(4\omega \pm 4\omega_3) \biggr \} \biggr ] \,,
 \nonumber \\
\frac{de^{do}}{d\tau} &= \frac{3 \pi}{16384} \frac{ \alpha \eta \epsilon^5 e (1-e^2)^{1/2}}{(1-E^2)^{7/2}} \biggl [
    4\sin^2 z \biggl \{ 2 {\cal G}_3 \sin 2 \omega +  e^2 (12212 + 53 E^2 +7200\sqrt{1-E^2}) \sin^2 z \sin 4 \omega \biggr \}
\nonumber \\
& \qquad
- 96 \cos z \sin^2 z (36-23e^2) \left (10+E^2 \right ) \sin 2\omega_3
\nonumber \\
& \qquad
+ 16 \left [ {\cal G}^\pm_4 (1 \mp \cos z)^2 \sin (2\omega \pm 2\omega_3 ) \right ]
- 8 e^2  {\cal G}_5 \sin^2 z  \left [ (1 \mp \cos z)^2 \sin (4\omega \pm 2\omega_3 ) \right ]
\nonumber \\
& \qquad
+ 3 E^2 \biggl \{ 2 (22-9e^2) \sin^2 z \left [ (1 \mp \cos z)^2 \sin (2\omega \pm 4\omega_3) \right ] + 13e^2 \left [ (1 \mp \cos z)^4 \sin (4\omega \pm 4\omega_3) \right ] \biggr \} \biggr ] \,,
 \nonumber \\
 \frac{d\iota^{do}}{d\tau} &= -\frac{3 \pi}{16384} \frac{ \alpha \eta \epsilon^5   \sin z}{(1-e^2)^{1/2} (1-E^2)^{7/2}} \biggl [
    4 e^2 \cos z \biggl \{ 2 {\cal G}_3 \sin 2 \omega +  e^2 (12212 + 53 E^2 +7200\sqrt{1-E^2}) \sin^2 z \sin 4 \omega \biggr \}
\nonumber \\
& \qquad
-16 {\cal G}_6 \sin 2\omega_3 
 +6 (8+72e^2-41e^4) E^2 \sin^2 z \sin 4\omega_3
 \nonumber \\
& \qquad
-8 e^2 \biggl \{ \left [2 {\cal G}^\pm_7 (\cos z \mp 1) \sin (2\omega \pm 2\omega_3) \right ] 
- e^2 \left [{\cal G}^\pm_8 (1 \mp \cos z )^2 \sin (4\omega \pm 2\omega_3) \right ]  \biggr \}
\nonumber \\
& \qquad
+3e^2 E^2 \biggl \{ 2(22-9e^2)  \left [ (1 \mp \cos z)^2 (\cos z \pm 2) \sin (2\omega \pm 4\omega_3) \right ]
    -13e^2   \left [ (\cos z \mp 1)^3 \sin (4\omega \pm 4\omega_3) \right ] \biggr \}  \biggr ]\,,
\nonumber \\ 
  \frac{d\Omega^{do}}{d\tau} &= \frac{3 \pi}{16384} \frac{ \alpha \eta \epsilon^5  }{(1-e^2)^{1/2} (1-E^2)^{7/2}} \frac{\sin z}{\sin \iota} \biggl [ 4 \cos z \biggl \{ {\cal G}_9  +4 e^2 {\cal G}_{10} \cos 2\omega \biggr \} 
\nonumber \\
& \qquad
+ 4 e^2 \cos z  \sin^2 z \left (12212 + 53E^2+ 7200 \sqrt{1-E^2} \right ) \cos 4 \omega
\nonumber \\
& \qquad
-2\cos z \biggl \{ 8{\cal G}_{11} \cos 2\omega_3 - 3 \sin^2 z (8+72e^2-41e^4) E^2 \cos 4 \omega_3 \biggr \}
\nonumber \\
& \qquad
+ 8 e^2 \biggl \{ 2\left [{\cal G}^\pm_{12} \,  (\cos z \mp 1)\cos (2\omega \pm 2\omega_3) \right ] 
+ e^2 \left [  {\cal G}^\pm_{8} (1 \mp \cos z)^2 \cos (4\omega \pm 2\omega_3) \right ] \biggr \}
\nonumber \\
& \qquad
+3 e^2 E^2 \biggl \{ 2(22-9e^2) \left [ (1 \mp \cos z)^2 (2\cos z \pm 1)  \cos (2\omega \pm 4\omega_3) \right ]
    -13 e^2 \left [  (\cos z \mp 1)^3   \cos (4\omega \pm 4\omega_3) \right ] \biggr \} \biggr ] \,.
\label{eq:QQtermsHex}
\end{align}
Note that $da/d\tau$ and $dA/d\tau$ are non-zero, but are {\em periodic} in the pericenter angles.  There are no ``purely secular'' terms, so these results do not violate the classic theorems.

The evolution of the inner and outer pericenters $\omega$ and $\omega_3$ are complicated by the fact that they also incorporate the evolution of $\Omega$, so we display them separately.  Including both dominant inner and outer feedback contributions and the outer time conversion contributions (labeled ``d''), they are given by 
\begin{align}
\frac{d\omega^{d}}{d\tau} &=-(\beta+\cos z ) \frac{\sin \iota}{\sin z}  \left (\frac{d\Omega^{di}}{d\tau} + \frac{d\Omega^{do}}{d\tau}\right )
\nonumber \\
& \quad +  \frac{3\pi}{128} \frac{\alpha^2 \epsilon^{9/2}}{(1+\alpha)^{1/2}} 
\frac{1}{(1-E^2)^3} \biggl [2 (3+2E^2) \cos z \biggl \{ (64-99e^2) + 3(12-17e^2) \cos^2 z 
\nonumber \\
& \qquad
+15(2-3e^2) \sin^2 z \cos 2\omega \biggr \}
+5E^2 H(E) \biggl \{ 5(2-3e^2) \left [  (1 \mp \cos z)^2 (3\cos z \pm 2) \cos(2\omega \pm 2\omega_3) \right ]
\nonumber \\
& \qquad
+6  \cos z \sin^2 z (17-12e^2) \cos 2\omega_3 \biggr \} \biggr ]\,.
 \nonumber \\
& \quad
 +\frac{\pi}{16384} \frac{ \alpha \eta \epsilon^5 (1-e^2)^{1/2}  }{(1-E^2)^{7/2}} \biggl [ 12 \biggl \{ {\cal G}_{13} +4 {\cal G}_{14} \sin^2 z \cos 2\omega + e^2 \sin^4 z \left (12212+53E^2+7200\sqrt{1-E^2} \right) \cos 4\omega \biggr \}
 \nonumber \\
& \qquad
+48 \sin^2 z \, {\cal G}_{15} \cos 2\omega_3 +18 E^2 (36-41e^2) \sin^4 z \cos 4\omega_3
 \nonumber \\
& \qquad
+ 24 \biggl \{ 4 \left [  {\cal G}^\pm_{16} (1 \mp \cos z )^2 \cos (2\omega \pm 2 \omega_3) \right ]
-  {\cal G}_{5} e^2 \sin^2 z \left [ (1 \mp \cos z)^2 \cos (4\omega \pm 2\omega_3) \right ] \biggr \}
\nonumber \\
& \qquad
+ 9 E^2 \biggl \{ 4 \sin^2 z (11-9e^2) \left [ (1 \mp \cos z)^2 \cos (2\omega \pm 4\omega_3) \right ]
 +13 e^2 \left [ (1 \mp \cos z)^4\cos (4\omega \pm 4\omega_3) \right ] \biggr \}
  \biggr ] \,,
\nonumber \\
\frac{d\omega_3^{d}}{d\tau} &= - (1+\beta \cos z) \frac{\sin \iota}{\sin z}  \left ( \frac{d\Omega^{di}}{d\tau}  + \frac{d\Omega^{do}}{d\tau}\right )
\nonumber \\
& \quad
+ \frac{3\pi}{64} \frac{\alpha^2 \beta \epsilon^{9/2}}{(1+\alpha)^{1/2}} 
\frac{1}{(1-E^2)^3} \biggl [ (22+8E^2) \cos z \biggl \{ 2 + 33e^2 - (2-17e^2) \cos^2 z +15 e^2 \sin^2 z \cos 2\omega \biggr \}
\nonumber \\
& \qquad 
+  G(E) \biggl \{ 5e^2 \left [ (1 \mp \cos z)^2 (3 \cos z \pm 2) \cos (2\omega \pm 2\omega_3) \right ]
- 2 \cos z \sin^2 z (2 -17e^2) \cos 2\omega_3 \biggr \} \biggr ] \,,
\nonumber \\
& \quad
+\frac{3\pi}{65536} \frac{ \alpha \eta \beta \epsilon^5}{ (1-e^2)^{1/2} (1-E^2)^{7/2}} \frac{1}{E^2}\biggl [4 \biggl \{  {\cal G}_{17} 
+4e^2 \sin^2 z \, {\cal G}_{18} \cos 2\omega 
+ e^4 \sin^4 z \,{\cal G}_{19} \cos 4\omega \biggr \}
\nonumber \\
& \qquad
+ 2 \biggl \{ 2\sin^2 z \, {\cal G}_{20} \cos 2\omega_3 
+3 (4 -7E^2 ) \sin^4 z (8 + 72e^2 -41e^4) \cos 4\omega_3 \biggr \}
\nonumber \\
& \qquad
-2e^2 \biggl \{ 4 \left [ {\cal G}^\pm_{21} (1 \mp \cos z)^2 \cos (2\omega \pm 2\omega_3) \right ] + e^2 \sin^2 z \,  {\cal G}_{22} \left [ (1 \mp \cos z)^2 \cos  (4\omega \pm 2\omega_3) \right ]
\nonumber \\
& \qquad
- 3e^2 (4-7E^2) \biggl \{ 4(22-9e^2) \sin^2 z \left [ (1 \mp \cos z)^2 \cos (2\omega \pm 4\omega_3) \right ] 
+13e^2 \left [ (1 \mp \cos z)^4 \cos (4\omega \pm 4\omega_3) \right ] \biggr \} \biggr ] \,,
\label{eq:QQdomomega}
\end{align}
where
\begin{align}
{\cal G}_{1} & = (64+16E^2-E^4) \left (3(8+72e^2-41e^4)\cos^2 z+ (24+88e^2-99e^4) \right ) \,,
\nonumber \\
{\cal G}^\pm_{2} &=  (22-9e^2) (1\mp \cos z)^2 (1 \pm 3\cos z +3\cos^2 z) \,,
\nonumber \\
{\cal G}_3 &= \left ((1382-1329 e^2 ) E^2 + 4(2582+471 e^2) \right ) \cos^2  z - (842+ 1641 e^2 ) E^2 - 4(442+1041e^2) 
\nonumber \\
& \qquad 
-480 (2+3e^2)(1-3\cos^2 z) \sqrt{1-E^2} \,,
\nonumber \\
{\cal G}^\pm_4 &= 160 (2+3 e^2) \left (1 \pm 5\cos z + 5\cos^2 z \right )  W(E)
- (8 (34+177 e^2) +15 (18+69 e^2) E^2) \cos^2 z 
\nonumber \\
& \qquad   
 \mp (2 (166 + 873 e^2)+ 12 (23+89e^2) E^2) \cos z   
 -4 (6+133 e^2)+ (66-127e^2) E^2  \,,
 \nonumber \\
 {\cal G}_5&= 1688+1305 E^2 - 4000 \,  W(E) \,,
\nonumber \\
 {\cal G}_6 &= 16 \left \{8 (1+3e^2 )(1-5 \cos^2 z) -(7+215 \cos^2 z) e^4 \right \}  W(E)
  \nonumber \\
& \qquad 
  +\left ((584+1416e^2+727e^4)+3(72+104e^2+483e^4) E^2 \right ) \cos^2 z 
  \nonumber \\
& \qquad 
- (280+984e^2-947e^4)- (88+312e^2+189e^4) E^2 \,,
 \nonumber \\
 {\cal G}^\pm_7 &= 160(2+3e^2) \left (1 \pm 5 \cos z +5\cos^2 z \right )  W(E)
   -\left (8(34+177e^2)+45(6+23e^2)E^2 \right ) \cos  z (\cos z \pm 1) 
  \nonumber \\
& \qquad 
   + 2(118-201e^2)+2(46-57e^2)E^2 \,,
  \nonumber \\ 
  {\cal G}^\pm_8 &= 2000  (2 \cos z \pm 1 ) W(E) -(1688+1305E^2)\cos z \mp (649+633E^2) \,,
\nonumber \\ 
{\cal G}_9 &= -\left (( 328 + 4872e^2 -5041e^4)E^2 + 4( 1288+ 7752e^2 + 119e^4) \right )\cos^2 z - (792+ 1638e^2 -7179e^4 )E^2  \nonumber \\
& \qquad 
-4(152 - 552e^2 -5619e^4 ) + 96\left ( 2(4+12e^2+9e^4)(1-3\cos^2 z)+75e^4 \sin^2 z \right ) \sqrt{1-E^2}\,,
\nonumber \\ 
{\cal G}_{10} &= \left ((1382 - 1329e^2  )E^2 +4(2582+471e^2) \right )\cos^2 z - 4(278 + 39e^2)E^2 - 3024(2+e^2) 
\nonumber \\
& \qquad
-480(2+3e^2)(2-3\cos^2 z)\sqrt{1-E^2}\,,
\nonumber \\ 
{\cal G}_{11} &= 32 \left \{ (8+24e^2+43e^4 )(3-5\cos^2 z) - 25e^4 \right \} W(E) 
+ \left (45 (8 + 8e^2 +71e^4 )E^2 + 8(56+24e^2+553e^4 ) \right )\cos^2 z
\nonumber \\
& \qquad 
 - (232 + 360e^2 + 1935e^4)E^2 - 2(72 - 120e^2 +1375e^4 ) \,,
\nonumber \\ 
{\cal G}^\pm_{12} &=80 (2+3 e^2 ) (3 \mp 5\cos z - 20 \cos^2 z )  W(E) + \left (16 (34+177 e^2 ) + 90 (6+23 e^2 ) E^2 \right )  \cos^2 z 
\nonumber \\
& \qquad 
\mp \left  ( (194- 843 e^2 ) - 3(34+177 e^2 ) E^2 \right ) \cos z - 21 (2 + 21e^2 ) - (194+ 417 e^2 ) E^2  \,,
\nonumber \\
{  \cal G}_{13} &= \left ((2436-5041  e^2 )  E^2 + 4  (3876+119  e^2 ) \right) \cos^4 z 
+ 2  \left ((684-7179  e^2 )  E^2 - 4  (276+5619  e^2 ) \right) \cos^2 z
\nonumber \\
& \qquad 
 + 3(556+ 549  e^2)  E^2 + 4 (2148+1367  e^2) -96 \left (8(2+3e^2)(2-3\cos^2 z)-3(12+43e^2) \sin^4 z \right )\sqrt{1-E^2} \,,
\nonumber \\
{\cal G}_{14} &= \left ((691-1329  e^2 )  E^2 + 4  (1291+471  e^2 ) \right )    \cos^2 z - (421+1641 e^2 )  E^2 - 4  (221+1041  e^2 ) 
\nonumber \\
& \qquad 
- 480(1+3e^2)(1-3\cos^2 z) \sqrt{1-E^2} \,,
\nonumber \\
{\cal G}_{15} &=32 \left ( 5(12+43  e^2 ) \cos^2 z   -(12-7  e^2 ) \right )   W(E) 
- \left ( 15 (12+213  e^2 )  E^2 + 8  (12-553  e^2 ) \right ) \cos^2 z  
\nonumber \\
& \qquad 
+  (268+ 477  e^2 )  E^2 + 8  (68-113  e^2 ) \,,
 \nonumber \\
{\cal G}^\pm_{16} &=160 (1+3e^2) (1 \pm 5 \cos z + 5 \cos^2 z )  W(E)
 - \left (45(3+23 e^2  ) E^2 + 8 (17 +177 e^2 ) \right )  \cos^2 z 
 \nonumber \\
& \qquad 
 \mp \left (3 (46+341 e^2 ) E^2 +2 (83+648 e^2 )\right )  \cos z
 + (33-97 e^2 ) E^2 - 4 (3+58 e^2 ) \,,
\nonumber \\
{\cal G}_{17} &=\left (( 1736 + 25224 e^2 -25697 e^4) E^4 + 12 (3592 + 23688 e^2- 3289 e^4 ) E^2 + 24 (344 + 3096 e^2-1763 e^4 ) \right)  \cos^4 z 
\nonumber \\
& \qquad 
+\left (6 (1416 + 2632 e^2 -12361 e^4 ) E^4 + 8 (3656 + 3144 e^2 -51657 e^4) E^2 + 48 (456 + 1288 e^2- 1809 e^4 ) \right )  \cos^2 z 
\nonumber \\
& \qquad 
+ (1736+ 17544 e^2 + 7743 e^4 ) E^4 + 4 ( 6296+ 43800 e^2 + 2645 e^4) E^2 + 24 (344 + 2584 e^2-1667 e^4 ) 
\nonumber \\
& \qquad
+192 \left ((8+24e^2+18e^4)(1-3\cos^2 z)^2 +225 e^4 \sin^4 z \right )E^2 \sqrt{1-E^2}
\,,
\nonumber \\
{\cal G}_{18} &= \left (( 7174 -6753 e^2) E^4 + 4 ( 23154 +837 e^2 ) E^2 + 24 (946 -387 e^2 ) \right )  \cos^2 z  
- (3946+8313 e^2 ) E^4 
\nonumber \\
& \qquad
- 4 (1502+8931 e^2) E^2 + 24 (418-171 e^2 )
-2880 (2+3e^2)(1-3\cos^2 z) E^2 \sqrt{1-E^2} \,,
\nonumber \\
{\cal G}_{19} &=421 E^4 + 95964 E^2 + 13416 + 43200 E^2 \sqrt{1-E^2} \,,
\nonumber \\
{\cal G}_{20} &=64 (1-3E^2) \left ((16+48e^2+111e^4)(1-5\cos^2 z)-125 e^4 \sin^2 z \right )E^2  Q(E) 
\nonumber \\
& \qquad 
- \left (3 (7480 + 17400 e^2 +49025 e^4  ) E^4 - 16 (248 +4632 e^2 - 5471 e^4) E^2 + 48 (24 + 216 e^2 -123 e^4 ) \right )  \cos^2 z 
\nonumber \\
& \qquad 
+ (6232 + 20376 e^2 +6357 e^4 ) E^4 + 16 (696 + 2520 e^2 -2215 e^4 ) E^2 - 48 (24 + 88 e^2 -99 e^4 ) \,,
\nonumber \\
{\cal G}^\pm_{21} &=640 (2+3 e^2 )(1 \pm 5 \cos z + 5 \cos^2 z) (1-3E^2) E^2  Q(E) 
\nonumber \\
& \qquad 
-  \left (15 (1618+3309 e^2 ) E^4 - 16 ( 1282 -1479 e^2 ) E^2 + 144 (22-9 e^2 ) \right )  \cos z ( \cos z \pm 1) 
\nonumber \\
& \qquad 
+ 5 (466+1581 e^2 ) E^4 - 48 (178-111 e^2 ) E^2 + 48 (22-9 e^2 ) \,,
\nonumber \\
{\cal G}_{22} &=16000 E^2 (1-3E^2 )  Q(E) + (1872 + 3152 E^2 +73905 E^4 ) \,,
\label{eq:Gcoeffs}
\end{align}
and
\begin{align}
W(E) &=  \frac{(1-E^2)}{1+\sqrt{1-E^2}} \,,
\nonumber \\
Q(E) &=  \frac{(1-E^2)}{(1+\sqrt{1-E^2})^2} \,,
\nonumber \\
E^2 H(E) &= \frac{1}{5} \left (2+3 E^2-4 W(E) \right ) \,,
\nonumber \\
G(E) &= 4+ 11E^2+ (2-5E^2) Q(E) 
\,,
\label{eq:QHG}
\end{align}

\subsection{Second-order perturbations: Subdominant $Q^2$ terms}

The subdominant  $Q^2$ terms from inner element feedback and inner time conversion (labeled ``s''), with amplitudes at $O(\alpha^2 \epsilon^6)$ are given by

\begin{align} 
\frac{da^s}{d\tau} &= -\frac{15\pi a}{2048} \frac{\alpha^2 \epsilon^6 e^2 (1-e^2)^{1/2}}{(1-E^2)^{9/2}}
\biggl [ 4(8+24E^2+3E^4) \sin^2 z \biggl \{ 2(1+9\cos^2 z) \sin 2\omega + 9 \sin^2 z \sin 4\omega \biggr \}
\nonumber \\
& \qquad
+16 E^2 (6+E^2) \biggl \{2\left [ {\cal K}^\pm_1 \sin (2\omega \pm 2\omega_3 ) \right ] + 3 \sin^2 z \left [ (1 \mp \cos z)^2 \sin (4\omega \pm 2 \omega_3) \right ] \biggr \}
\nonumber \\
& \qquad
+ 3E^4 \biggl \{ 2 \sin^2 z \left [ (1 \mp \cos z)^2 \sin (2\omega \pm 4 \omega_3 ) \right ] + \left [(1 \mp \cos z)^4 \sin (4\omega \pm 4\omega_3 ) \right ] \biggr \}  \biggr ] \,,
\nonumber \\
\frac{dA^s}{d\tau} &=-\frac{45\pi A}{1024} \frac{\alpha^2 \beta \epsilon^6 (1-e^2)^{1/2}}{(1-E^2)^{11/2}}
\biggl [ 8 \sin^2 z \cos z (16 +120E^2 +90E^4 + 5E^6 ) \sin 2 \omega 
\nonumber \\
& \qquad + 5E^2 (16+16E^2 + E^4) \left [ (1 \mp \cos z)^2 (3\cos z \pm  2) \sin (2\omega \pm 2\omega_3) \right ] \biggr ] \,,
\nonumber \\
\frac{de^s}{d\tau} &= \frac{3\pi}{16384} \frac{\alpha^2 \epsilon^6 e(1-e^2)^{1/2}}{(1-E^2)^{9/2}}
\biggl [ 4\sin^2 z \biggl \{ 2 {\cal K}_2 \sin 2\omega - 15 {\cal K}_3 \sin^2 z \sin 4\omega \biggr \}
\nonumber \\
& \qquad
+16 E^2 (6+E^2) \biggl \{2 \left [ {\cal K}^\pm_4 \sin (2\omega \pm 2\omega_3) \right ]
+5 (78 + 83e^2) \sin^2 z \left [ (1 \mp \cos z)^2 \sin (4\omega \pm 2\omega_3 ) \right ] \biggr \}
\nonumber \\
& \qquad
+ 5E^4 \biggl \{ 2  (150+11e^2) \sin^2 z \left [  (1 \mp \cos z)^2 \sin (2\omega \pm 4\omega_3 ) \right ] + (78 + 83e^2) \left [  (1 \mp \cos z)^4 \sin (4\omega \pm 4\omega_3 ) \right ] \biggr \}
\biggr ] \,,
\nonumber \\
\frac{d\iota^s}{d\tau} &= -\frac{3\pi}{16384} \frac{\alpha^2 \epsilon^6 }{(1-e^2)^{1/2}(1-E^2)^{9/2}} \sin z
\biggl [
4 \cos z \biggl \{2 {\cal K}_5 \sin 2 \omega - 15 e^2 {\cal K}_6 \sin^2 z \sin 4 \omega \biggr \}
\nonumber \\
& \qquad
+2 E^2  \biggl \{ 8 (6+E^2) {\cal K}_7 \sin 2 \omega_3 + 5 E^2 (56+494e^2-67e^4) \sin^2 z \sin 4 \omega_3  \biggr \}
\nonumber \\
& \qquad
-8 E^2 (6+E^2) \biggl \{4 \left [ {\cal K}^\pm_8 \sin (2\omega \pm 2\omega_3) \right ]
-5 e^2 (90 + 71e^2) \left [ (1 \mp \cos z)^2 (2\cos z \pm 1) \sin (4\omega \pm 2\omega_3 ) \right ] \biggr \}
\nonumber \\
& \qquad
+ 5 E^4  \biggl \{ 2 \left [{\cal K}^\pm_{9} \sin (2\omega \pm 4\omega_3 ) \right ]
-e^2 (90+71e^2 ) \left [ ( \cos z \mp 1)^3  \sin (4\omega \pm 4\omega_3 ) \right ] \biggr \}
\biggr ] \,,
\nonumber \\
\frac{d\Omega^s}{d\tau} &=\frac{3\pi}{16384} \frac{\alpha^2 \epsilon^6 }{(1-e^2)^{1/2}(1-E^2)^{9/2}} 
\frac{\sin z}{\sin \iota} \biggl [
4 \cos z \biggl \{ {\cal K}_{10} - 4 {\cal K}_{11}  \cos 2 \omega - 15 e^2 {\cal K}_{6} \sin^2 z \cos 4 \omega \biggr \}
\nonumber \\
& \qquad
+2 E^2 \cos z \biggl \{ 16 (6+E^2) {\cal K}_{12} \cos 2 \omega_3 + 5 E^2 (56+494e^2-67e^4) \sin^2 z \cos 4 \omega_3  \biggr \}
\nonumber \\
& \qquad
-8 E^2 (6+E^2) \biggl \{2 \left [ {\cal K}^\pm_{13} \cos (2\omega \pm 2\omega_3) \right ]
-5 e^2 (90 + 71e^2)  \left [ (1 \mp \cos z)^2 (2\cos z \pm 1) \cos (4\omega \pm 2\omega_3 ) \right ] \biggr \}
\nonumber \\
& \qquad
+ 5 E^4  \biggl \{ 2 \left [{\cal K}^\pm_{14} \cos (2\omega \pm 4\omega_3 ) \right ]
-e^2 (90+71e^2 ) \left [ ( \cos z \mp 1)^3  \cos (4\omega \pm 4\omega_3 ) \right ] \biggr \}
\biggr ] \,,
\nonumber \\
\frac{d\omega^s}{d\tau} &= -\cos z \frac{\sin \iota}{\sin z} \frac{d\Omega^s}{d\tau} - \frac{\pi}{8192} \frac{\alpha^2 \epsilon^6 (1-e^2)^{1/2}}{(1-E^2)^{9/2}}
\biggl [
12 \biggl \{  {\cal K}_{15} -2  {\cal K}_{16} \sin^2 z \cos 2\omega +15 {\cal K}_{17}  \sin^4 z \cos 4 \omega  \biggr \}
\nonumber \\
& \qquad
-30 E^2 \sin^2 z  \biggl \{ 16(6+E^2) \left ((155-32e^2) \cos^2 z +103-100e^2 \right ) \cos 2 \omega_3
+ E^2  \sin^2 z  (155-32e^2) \cos 4\omega_3 \biggr \}
\nonumber \\
& \qquad
-48 E^2 (6+E^2) \biggl \{ 2 \left [(1 \mp \cos z)^2 {\cal K}^\pm_{18} \cos (2\omega \pm 2\omega_3 ) \right ]
+5  \sin^2 z (39+34e^2) \left [(1 \mp \cos z)^2 \cos (4\omega \pm 2\omega_3 ) \right ] \biggr \}
\nonumber \\
& \qquad
-15 E^4 \biggl \{ 2 (99-e^2) \sin^2 z \left [(1 \mp \cos z)^2  \cos (2\omega \pm 4\omega_3 ) \right ]
+ (39+34e^2) \left [(1 \mp \cos z)^4  \cos (4\omega \pm 4\omega_3 ) \right ] \biggr \}
\biggr ] \,,
\nonumber \\
\frac{d\omega_3^s}{d\tau} &= -(1+\beta \cos z) \frac{\sin \iota}{\sin z} \frac{d\Omega^s}{d\tau} 
\nonumber \\
& \quad +\frac{\pi}{65536} \frac{\alpha^2 \beta \epsilon^6 }{(1-e^2)^{1/2}(1-E^2)^{9/2}}
\biggl [
4 \biggl \{  {\cal K}_{19} + 12  {\cal K}_{20}  \cos 2\omega - 135 {\cal K}_{21} e^2 \sin^4 z \cos 4 \omega  \biggr \}
\nonumber \\
& \qquad
- 6 \biggl \{ 4 {\cal K}_{22} \cos 2 \omega_3
+ E^2 \sin^2 z {\cal K}_{23} \cos 4\omega_3 \biggr \}
+12  \biggl \{ \left [ (\cos z \mp 1) {\cal K}^\pm_{24} \cos (2\omega \pm 2\omega_3 ) \right ]
\nonumber \\
& \qquad
+20e^2 (90+71e^2) (12+46E^2+5E^4) \sin^2 z \left [ (1 \mp \cos z)^2  \cos (4\omega \pm 2\omega_3 ) \right ] \biggr \}
\nonumber \\
& \qquad
+3 E^2  \biggl \{ 4  {\cal K}_{25}  \left [ (1 \mp \cos z)^2  \cos (2\omega \pm 4\omega_3 ) \right ]
+5e^2 (90+71e^2)(4+5E^2) \left [ (1 \mp  \cos z )^4  \cos (4\omega \pm 4\omega_3 ) \right ] \biggr \}
\biggr ] \,,
 \label{QQtermsSubdom}
\end{align}
where
\begin{align}
{\cal K}^\pm_1 &= (1 \mp \cos z)^2 (1 \pm 3 \cos z +3\cos^2 z ) \,,
\nonumber \\
{\cal K}_2 &=  320 (1-E^2)^{3/2} (7+3e^2 )(1-3\cos^2 z) + 3(8+24E^2+3E^4) \left (5(150+11e^2)\cos^2 z + 206-17e^2 \right )\,,
\nonumber \\
{\cal K}_3 &= 320 (1-E^2)^{3/2} (1+e^2) - (8+24E^2+3E^4) (78+83e^2) \,,
\nonumber \\
{\cal K}^\pm_4 &= (1 \mp \cos z)^2 \left ( 5 (150+11e^2 )\cos z (\cos z \pm 1) + 342 + e^2 \right ) \,,
\nonumber \\ 
{\cal K}_5 &=  e^2 {\cal K}_2  +20 (8+24E^2+3E^4) (1-e^2) (12 + e^2 +9e^2 \cos^2 z) \,,
\nonumber \\
{\cal K}_6 &= {\cal K}_3 - 12 (1-e^2) (8+24E^2+3E^4)  \,,
\nonumber \\
{\cal K}_7 &= \left (5( 56 + 494e^2 -67e^4) \cos^2 z - (328-1910e^2 + 1015e^4) \right ) \,,
\nonumber \\
{\cal K}_8^\pm &= (\cos z \mp 1)\left (5e^2 (162-e^2)\cos^2 z \pm 5(12+150e^2-e^4) \cos z +(60+302e^2 -19e^4) \right ) \,,
\nonumber \\
{\cal K}_9^\pm &= (1 \mp \cos z)^2 \left ( e^2 (162-e^2)\cos z \pm 2(12+150e^2 -e^4) \right ) \,,
\nonumber \\
{\cal K}_{10} &= 64  (1-E^2)^{3/2} \left ( 3 (28+79e^2+43e^4) \cos^2 z -(28+129e^2+93e^4) \right ) 
\nonumber \\
& \qquad
        -3(8+24E^2+3E^4)\left ( 5(56+494e^2-67e^4) \cos^2 z - (312-450e^2+565e^4) \right ) \,,
\nonumber \\
{\cal K}_{11}&= 320 (1-E^2)^{3/2} \left (3(1+6e^2+3e^4) \cos^2 z - (3+11e^2+6e^4) \right ) 
\nonumber \\
& \qquad
       - (8+24E^2+3E^4) \left (15 (12+150e^2 -e^4) \cos^2 z  -4(15+209e^2 +7e^4) \right ) \,,
\nonumber \\
{\cal K}_{12}&= \left (5(56+494e^2-67e^4) \cos^2 z - 4(76+70e^2+85e^4)\right ) \,,
\nonumber \\
{\cal K}^\pm_{13}&=(\cos z \mp 1)(20(12+150e^2-e^4)\cos^2 z \pm 5e^2(162-e^2)\cos z -e^2(86+33e^2)) \,,
\nonumber \\
{\cal K}^\pm_{14}&=(1 \mp \cos z)^2 \left (2(12+150e^2-e^4)\cos z \pm e^2(162-e^2) \right )  \,,
\nonumber \\
{\cal K}_{15}&=32 (1-E^2)^{3/2} \left ( 89+81e^2-6(39+31e^2) \cos^2 z +3(67+43e^2) \cos^4 z \right ) 
\nonumber \\
& \quad 
-5(8+24E^2+3E^4) \left (577-336e^2+6(17-56e^2)\cos^2 z +3(155-32e^2) \cos^4 z  \right ) \,,
\nonumber \\
{\cal K}_{16}&= 320 (1-E^2)^{3/2} (5+3e^2)(1-3\cos^2 z)+(8+24E^2+3E^4) \left (15(99-e^2)\cos^2 z +349 -71e^2 \right )\,,
\nonumber \\
{\cal K}_{17}&=  160 (1-E^2)^{3/2} (1+e^2) - (8+24E^2+3E^4) (39+34e^2) \,, 
\nonumber \\
{\cal K}^\pm_{18} &= 5(99-e^2) \cos z (\cos z \pm 1) + 211 - 19e^2 \,,
\nonumber \\
{\cal K}_{19}&=-384 (1-E^2)^{3/2} \left ( 9(28+79e^2+43e^4) \cos^4 z - 6(28+129e^2+93e^4) \cos^2 z + (28+279e^2+243e^4) \right ) 
\nonumber \\
& \quad
+675( 8+ 12E^2 +E^4 )(56 + 494e^2 -67e^4 )\cos^4 z - 18 \left (( 4152- 14230e^2 +11555e^4 )E^4 
\right .
\nonumber \\
& \quad \quad
\left .
+ 12( 4188- 13720e^2 +11345e^4 )E^2 + 8(4224 - 13210e^2 +11135e^4 ) \right )\cos^2 z 
\nonumber \\
& \quad
+ 3( 5336+ 169310e^2  -57235e^4)E^4 + 36( 5264+ 168290e^2  -56815e^4)E^2 + 24( 5192 + 167270e^2 -56395e^4) \,,
\nonumber \\
{\cal K}_{20}&= 1920 (1-E^2)^{3/2} (1+6e^2+3e^4) \sin^2 z (1-3\cos^2 z)
- 225 ( 8+ 12 E^2 +  E^4) ( 12+ 150 e^2 -e^4) \cos^4 z 
\nonumber \\
& \quad \quad
+ 4 \left (3 ( 200+ 1281 e^2 +213 e^4 ) E^4 
+ 9 ( 800+ 5331 e^2 + 813 e^4) E^2 
+ 12 (400 + 2769 e^2 +387 e^4 )\right ) \cos^2 z 
\nonumber \\
& \quad \quad
 + ( 300+ 12306 e^2 -1637 e^4 ) E^4
+ 12 ( 300+ 12099 e^2  -1598 e^4) E^2 + 8 ( 300 + 11892 e^2-1559 e^4  )  \,,
\nonumber \\
{\cal K}_{21}&= 640 (1-E^2)^{3/2}(1+e^2)  - 5 ( 8 + 12 E^2 + E^4) (90+71e^2) \,,
\nonumber \\
{\cal K}_{22}&=20 ( 56+ 494 e^2 -67 e^4 ) ( 12+ 46 E^2 +5 E^4 ) \cos^4 z
 - \left (( 11122-3505 e^2 +19655 e^4 ) E^4 
 \right .
\nonumber \\
& \quad \quad
\left .
+ 16 (6434 -1465 e^2 +11075 e^4 ) E^2 
+ 16 (1674 - 445 e^2 +2915 e^4 ) \right ) \cos^2 z 
\nonumber \\
& \quad \quad 
+ 5 (1222 - 8915 e^2 +4585 e^4 ) E^4 + 8 (6932 - 52600 e^2 + 26915 e^4) E^2 + 48 (274 - 2675 e^2 +1330 e^4 ) \,,
\nonumber \\
{\cal K}_{23} &= 5 (56 + 494 e^2 -67 e^4 ) (4+5 E^2 ) \cos^2 z - ( 1448+ 13030 e^2 -1955 e^4 ) E^2 - 4 ( 328+ 3150 e^2 -615 e^4 ) \,,
\nonumber \\
{\cal K}^\pm_{24} &= 80 ( 12+ 150 e^2 -e^4 ) ( 12+ 46 E^2 + 5 E^4) \cos^3 z \mp 120 (1- e^2 ) (144+112E^2+5E^4) \cos^2 z 
\nonumber \\
& \quad \quad
- \left ((3000 + 20903 e^2 +3369 e^4 ) E^4 + 16 ( 1560+ 12631 e^2 +1853 e^4 ) E^2 
+ 16 (120 + 3531 e^2 +493 e^4 ) \right) \cos z 
\nonumber \\
& \quad \quad 
\mp 5 (  240+ 5207 e^2 -239 e^4) E^4 \mp 16 ( 360+ 15119 e^2 -653 e^4 ) E^2 \pm 16 (480 - 4203 e^2 +111 e^4 ) \,,
\nonumber \\
{\cal K}_{25} &=5 (4+5 E^2 ) (12+ 150 e^2 -e^4  ) \sin^2 z  +2(4+E^2) e^2 (69-13e^2) \,,
\label{eq:Kcoeffs}
 \end{align}
Here again, $da/d\tau$ and $dA/d\tau$ are non-zero, but are periodic in the pericenter angles.
\subsection{Second-order perturbations: Dominant Quadrupole-Octopole cross-terms}

These terms arise from the inner and outer element feedback of octopole terms into the quadrupole perturbations and vice versa, as well as the quadrupole perturbations corrected by the outer octopole time conversion and vice versa.  The formulae are too long to be displayed here; instead we have made the raw formulae for all the results of this appendix available at \url{github.com/landenconway/Three-Body-Secular-Equations}.

The dominant quadrupole-octopole terms from inner element feedback have amplitudes proportional to $\alpha^2 (1+\alpha)^{-1/2} \Delta \epsilon^{11/2}$ for the inner elements as expected (see Tables  \ref{tab:table1} and \ref{tab:table2}) , and the same amplitude multiplied by $\beta$ for the outer elements.  Interestingly, for these terms, both $a$ and $A$ are constant.   The dominant quadrupole-octopole terms from outer element feedback have amplitudes proportional to $\alpha \eta \Delta \epsilon^{6}$ for the inner elements as expected, and the same amplitudes multiplied by $\beta$ for the outer elements.  However, for these terms, both $da/d\tau$ and $dA/d\tau$ are non-zero.

 \section{Constructing the orbit element perturbation $Y_\beta^{(0)}(t)$ in the secular approximation}
 \label{app:Y}
 
Here we show that the orbit element perturbation  $Y_\beta^{(0)}(t)$, Eq.\ (\ref{eq:Ydef}), which involves an integral over time, can actually be expressed as a sum of products of functions of the form $A(t) \times M(t)$, where $A(t)$ is periodic on the inner orbital timescale and $M(t)$ is periodic on the outer orbital timescale.    We assume as usual that $Q_\beta^{(0)}$ 
is a sum of functions of the same form.   We break the integral $\int_0^t A(t')M(t') dt'$ into subintervals of size $P_1$ and expand the function $M(t') = M_r + (t' - rP_1) \dot{M}_r + \dots$, where $M_r = M(rP_1)$, and we recall that $\dot{M} \sim M/P_2$.  The result is
\begin{align}
\int_0^t A(t')M(t') dt' &= \sum_{r=0}^{q-1} \int_{rP_1}^{(r+1)P_1} A(t') \left [ M_r + (t' - rP_1) \dot{M}_r + \dots \right ] dt'
+\int_{qP_1}^{t} A(t') dt' M_q + O \left (\frac{P_1^2}{P_2} AM \right) 
\nonumber \\
&= \sum_{r=0}^{q-1} P_1 \left ( \langle A \rangle  M_r + \langle tA \rangle \dot{M}_r \right ) 
+ M(t) \int_{qP_1}^{t} A(t') dt'  + O(P_1^2/P_2 \times AM) \,.
\label{eq:Y1}
\end{align}
Using the fact that $\int_0^t M(t') dt' =   \sum_{r=0}^{q-1} \left ( P_1M_r +  P_1^2 \dot{M}_r \right ) + (t-qP_1) M_q +O(MP_1^2/P_2)$, we can rewrite Eq.\ (\ref{eq:Y1}) in the form
 \begin{align}
\int_0^t A(t')M(t') dt' &= \langle A \rangle \int_0^t M(t') dt' + M(t) \int_0^t {\cal AF} \left (A(t') \right ) dt'
 + \left \langle \left (t - P_1/2 \right ) A \right \rangle \left ( M(t) - M(0) \right )  \,.
\end{align}
From Eq.\ (B13) of \cite{2021PhRvD.103f3003W}, we can show that
\begin{equation}
\left \langle \int_0^t A(t')M(t') dt' \right \rangle = \langle A \rangle \left \langle \int_0^t M(t') dt' \right \rangle - \langle \left ( t- P_1/2 \right ) A \rangle M(0) \,.
\end{equation}
Pulling everything together, we obtain the main ingredient for constructing $Y_\beta^{(0)}(t)$:
\begin{equation}
{\cal AF} \int_0^t {\cal AF} \left (A(t') M(t') \right) dt' = \langle A \rangle {\cal AF} \int_0^t {\cal AF} \left ( M(t') \right ) dt'
+  M(t) {\cal AF} \int_0^t {\cal AF} \left (A(t') \right) dt' + O \left ( \frac{P_1^2}{P_2} AM \right ) \,.
\label{eq:Yfinal}
\end{equation}
The first term is the dominant term, of order $P_2 AM$, while the second is the subdominant term, of order $P_1 AM$; both terms are of the form of a product of functions of the inner and outer orbit timescales respectively.   This property will prove useful in applying the secular approximation at second order in perturbation theory.  In terms of explicit integrals over $u$ and $F$, the dominant and subdominant contributions to $Y_\beta^{(0)}$ take the forms
\begin{align}
Y_\beta^{(0) do} & = \int_0^F \left [ \frac{\tilde{n}}{2\pi} \int_0^{2\pi} \widehat{Q}^{(0)}_\beta du - \frac{dt}{dF} \left \langle {Q}^{(0)}_\beta \right \rangle \right ] dF'
- \frac{\tilde{N}}{2\pi} \int_0^{2\pi} \left \{ \dots \right \} \frac{dt}{dF}  dF \,,
\nonumber \\
Y_\beta^{(0) s} & = \int_0^u \left [ {Q}^{(0)}_\beta - \frac{\tilde{n}}{2\pi} \int_0^{2\pi} {Q}^{(0)}_\beta \frac{dt}{du'} du' \right ] \frac{dt}{du}  du 
- \frac{\tilde{n}}{2\pi} \int_0^{2\pi} \left \{ \dots \right \} \frac{dt}{du}  du \,,
\end{align} 
where the quantity $\{ \dots \}$ denotes the indefinite integral to the left.

 \section{Corrections arising from the $dt/du$ and $dt/dF$ conversions}
 \label{app:timeconversion}
 
 In Eq.\ (\ref{eq:T1T2}), we must evaluate the term 
 \begin{align}
\int_0^{2\pi}  \int_0^{2\pi}  \left [ \left (\frac{dt}{du} \right )^{(0)}_{K,\beta}  \left (\frac{dt}{dF} \right )^{(0)}_K +
\left (\frac{dt}{du} \right )^{(0)}_K  \left (\frac{dt}{dF} \right )^{(0)}_{K,\beta} \right ] Y^{(0)}_\beta du dF \,. 
 \end{align}
Substituting for  ${dt/du}^{(0)}_K$ in the first term, we obtain
\begin{equation}
\int_0^{2\pi}  \int_0^{2\pi} \left [ - \frac{\cos u}{\tilde{n}} \left (\frac{dt}{dF} \right )^{(0)}_K  Y^{(0)}_e
+ \frac{3}{2a} \left (\frac{dt}{du} \right )^{(0)}_{K}  \left (\frac{dt}{dF} \right )^{(0)}_K Y^{(0)}_a \right ] du dF \,.
\end{equation}
The second term is proportional to $\langle Y^{(0)}_a \rangle = 0$, while the first term can be integrated by parts to yield
\begin{align}
\int_0^{2\pi}  \int_0^{2\pi} &\frac{\sin u}{\tilde{n}} \left (\frac{dt}{dF} \right )^{(0)}_K \frac{\partial}{\partial u} \left ( Y^{(0) do}_e + Y^{(0) s}_e \right ) du dF 
 = \frac{2\pi}{\tilde{n}^2}\frac{2\pi}{\tilde{N}} \left \langle \sin u Q_e^{(0)} \right \rangle  \,,
\end{align}
where we use the fact that $\partial Y^{(0) do}_e /\partial u = 0$.  Substituting for ${dt/dF}^{(0)}_K$ in the second term, we obtain
\begin{equation}
\int_0^{2\pi}  \int_0^{2\pi} \left [  \frac{1}{\tilde{N}} \frac{\partial}{\partial \tilde{E}} \left ( \frac{(1-\tilde{E}^2)^{3/2}}{(1+\tilde{E}\cos F)^2} \right ) \left (\frac{dt}{du} \right )^{(0)}_K  Y^{(0)}_E
+ \frac{3}{2A} \left (\frac{dt}{du} \right )^{(0)}_{K}  \left (\frac{dt}{dF} \right )^{(0)}_K Y^{(0)}_A \right ] du dF \,.
\end{equation}
Noting that the second term is proportional to $\langle Y^{(0)}_A \rangle = 0$, and that 
\begin{equation}
\frac{\partial}{\partial \tilde{E}} \left ( \frac{(1-\tilde{E}^2)^{3/2}}{(1+\tilde{E}\cos F)^2} \right ) = \sqrt{1-E^2} \frac{\partial}{\partial F} \left ( \frac{\sin F (2+ \tilde{E} \cos F)}{(1+\tilde{E}\cos F)^2} \right ) \,,
\end{equation}
we integrate by parts with respect to $F$, and arrive finally at the result
\begin{equation}
 \frac{2\pi}{\tilde{n}}\frac{2\pi}{\tilde{N}^2} \sqrt{1-\tilde{E}^2} \left \langle  \frac{\sin F (2+ \tilde{E} \cos F)}{(1+\tilde{E}\cos F)^2}  Q_E^{(0)} \right \rangle \,.
\end{equation}
These results appear in Eq.\ (\ref{eq:T1T2}).
 
  \end{widetext}
 \section{Constancy of the semimajor axes at first order}
 \label{app:semimajor}
 
In this appendix we provide a simple proof that both the inner and outer semimajor axes $a$ and $A$ are constant to first order in perturbation theory and to arbitrary multipole orders.  The Lagrange planetary equation for the semimajor axis is given by suitably combining the equations for $dp/dt$ and $de/dt$ in Eq.\ (\ref{eq2:lagrange}), leading to
\begin{equation}
\frac{da}{dt} = \frac{2a}{1-e^2} \sqrt{\frac{p}{Gm}} \biggl [ e \sin f {\cal R} + (1+e \cos f) {\cal S} \biggr ] \,,
\label{eq:dadt}
\end{equation}
with an analogous equation for $dA/dt$.  
For a chosen value of multipole index $\ell$, we 
substitute the perturbing acceleration for the inner orbit in Eq.\ (\ref{eq2:eom2}) into  Eq.\ (\ref{eq:dadt}), to obtain
\begin{align}
\frac{da}{dt} &= \frac{2a}{1-e^2} \sqrt{\frac{p}{Gm}} \, \frac{Gm_3 p^\ell}{R^{\ell +2}} \,\frac{(2\ell +1)!!}{\ell!} \frac{g_\ell}{(1+e \cos f)^\ell}
\nonumber \\
& \times \left [ e \sin f \, n^{jL}  + (1+e \cos f) \lambda^j n^L \right ] N^{\langle jL \rangle }  \,.
\end{align}
From the properties of the symmetric trace-free tensor products $N^{\langle L \rangle }$ (see Eqs.\ (1.159) and (1.160) of \cite{PW2014}), we have that
\begin{equation}
n^{L} N^{\langle jL \rangle } =  \frac{\ell!}{(2\ell+1)!!} \left ( N^j  \frac{dP_{\ell +1} (\zeta)}{d\zeta} - n^j \frac{dP_{\ell} (\zeta)}{d\zeta} \right ) \,,
\end{equation}
where $P_\ell$ is the Legendre polynomial with argument $\zeta = \bm{n} \cdot \bm{N}$, satisfying the recursion relation
$\zeta dP_{\ell +1}/d\zeta - dP_{\ell}/d\zeta = (\ell + 1) P_{\ell +1}$.    Noting that ${\bm \lambda} = \partial {\bm n}/\partial f$ and thus that ${\bm \lambda} \cdot {\bm N} = \partial \zeta/\partial f$, we can show that 
\begin{align}
& \frac{ \left [ e \sin f \, n^{jL}  + (1+e \cos f) \lambda^j n^L \right ] N^{\langle jL \rangle} }{(1+e \cos f)^\ell} 
\nonumber \\
& \qquad
= (1+e \cos f)^2 \frac{\partial}{\partial f} \left ( \frac{P_{\ell +1} (\zeta)}{(1+e \cos f)^{\ell + 1}} \right ) \,.
\label{eq:simplify}
\end{align}
Then in the secular approximation at first order in the perturbations, the average of $da/dt$ is given by
\begin{align}
\left \langle \frac{da}{dt} \right \rangle &= \frac{\tilde{n}\tilde{N}}{(2\pi)^2} \int_0^{2\pi} \biggl (\frac{dt}{dF} \biggr )_K dF  \int_0^{2\pi}  \biggl (\frac{dt}{df}  \biggr)_K \frac{da}{dt} df \,,
\label{eq:avdadt}
\end{align}
holding the orbit elements fixed.  Here $\tilde{n}$ and $\tilde{N}$ are the mean anomalies.
Substituting $(dt/df)_K = (p^3/Gm)^{1/2} (1+e \cos f)^{-2}$ and Eq.\ (\ref{eq:simplify}) into (\ref{eq:avdadt}), and integrating over $f$ holding $F$ fixed, we see that the $f$ integration yields $[P_{\ell + 1}(\zeta)/(1+e \cos f)^{\ell +1}]_0^{2\pi} = 0$, for any $\ell$.

In the same manner, substituting  the perturbing acceleration for the outer orbit in Eq.\ (\ref{eq2:eom2}) into the equation for $dA/dt$, expressing the results in terms of Legendre polynomials, and using the fact that ${\bm \Lambda} \cdot {\bm n} = \partial \zeta /\partial F$, we can convert the $F$ integral in the secular approximation into  $[P_{\ell + 1}(\zeta)(1+E \cos F)^{\ell +2}]_0^{2\pi} = 0$.
Thus, at first order in perturbation theory, $\langle da/dt \rangle$ and $\langle dA/dt \rangle$ both vanish to all multipole orders.


\end{document}